\documentstyle[preprint,tighten,aps]{revtex}

\newcommand{\delv}{{\bf \nabla}}
\newcommand{\delvc}{{\bf D}}

\newcommand{\delfour}{{\Delta^{(4)}}}
\newcommand{\delsq}{\Delta^{(2)}}

\newcommand{\vev}[1]{\langle #1 \rangle}

\newcommand{\Sigmav}{\mbox{\boldmath$\Sigma$}}
\newcommand{\gammav}{\mbox{\boldmath$\gamma$}}
\newcommand{\alphav}{\mbox{\boldmath$\alpha$}}

\newcommand{\fbs}{{$f_{B_s}\,$}}
\newcommand{\fds}{{$f_{D_s}\,$}}
\newcommand{\Mbz}{{M_0}}

\newcommand{\be}{\begin{equation}}
\newcommand{\ee}{\end{equation}}

\newcommand{\lag}{{\cal L}}

\newcommand{\nl}{\nonumber \\}

\newcommand{\Ev}{{\bf E}}
\newcommand{\Bv}{{\bf B}}

\newcommand{\sigmav}{\mbox{\boldmath$\sigma$}}
\newcommand{\ainv}{$a^{-1}$}
\newcommand{\ainvsp}{$a^{-1}\;$}

\begin{document}
\preprint{\begin{minipage}{2in}\begin{flushright}
 OHSTPY-HEP-T-00-004 \\ GUPTA/00-06-06 \\ CLNS 00-1678
\end{flushright}
  \end{minipage}}
\input epsf
\draft

\title{ {\bf Scaling and Further Tests of Heavy Meson  \\
Decay Constant Determinations from NRQCD}}

\author{
  S.Collins$^{a*}$, C.T.H.Davies$^a$\footnote{UKQCD Collaboration},
 J.Hein$^b$, G.P.Lepage$^b$,\\
 C.J.Morningstar$^c$, J.Shigemitsu$^d$, 
J.Sloan$^{e}$\footnote{Present address: Spatial Technologies, 
Boulder, CO, USA.}\\[.4cm]
\small $^a$Department of Physics \& Astronomy, University of Glasgow, 
 Glasgow, G12 8QQ, UK. \\[.2cm]
\small $^b$Newman Laboratory of Nuclear Studies, Cornell University, 
 Ithaca, NY 14853, USA. \\[.2cm]
\small $^c$Department of Physics, Florida International University,
 Miami, FL 33199, USA  \\[.2cm]
\small $^d$Physics Department, The Ohio State University,
 Columbus, OH 43210, USA.\\[.2cm]
\small $^e$Physics Department, University of Kentucky,
 Lexington, KY 40506, USA.
\\ }


\maketitle

\begin{abstract}
\noindent

We present results for the $B_s$ meson decay constant \fbs from simulations 
at three lattice spacings in the range \ainv = 1.1GeV to 2.6GeV 
using NRQCD heavy quarks and clover light quarks in the quenched 
approximation.  We study scaling of this 
quantity and check consistency between mesons decaying from rest and from 
a state with nonzero spatial momentum.  Cancellation of power law contributions
 that arise in the NRQCD formulation of heavy-light currents is discussed.  
On the coarsest lattice the $D_s$ meson decay constant $f_{D_s}$ is 
calculated.  Our best values for the decay constants are given by 
$f_{B_s} = 187(4)(4)(11)(2)(7)(6) \; {\rm MeV}$ and 
$f_{D_s} = 223(6)(31)(38)(23)(9)(^{+3}_{ -1})\; {\rm MeV}$.

\end{abstract}

\vspace{.2in}

\newpage

\section{Introduction}

The $B$ meson decay constant, $f_B$, defined through the 
hadronic matrix element
\be \label{deffb}
   \langle \, 0 \,| \, A_{\mu} \,|\,B,\,\vec{p} \, \rangle = p_{\mu} 
f_{B} ,
\ee
is one of the phenomenologically relevant quantities whose determination 
 has relied heavily on lattice calculations.  Together with the bag-parameter 
$B_B$, the decay constant is an important ingredient in analyses of 
$B^0\overline{B^0}$ mixing phenomena and studies of CP violation in the 
neutral $B$ system.  Several quenched
 lattice determinations of $f_B$ have been 
completed in recent years using different approaches to simulating the 
heavy $b$-quark on the lattice 
\cite{fb,jlqcd98,fbplb,jlqcdhiro,fermifb,milc,ape,ukqcd}. 
 The first dynamical results are also starting to appear 
\cite{sara,cppacs1,cppacs2,milc}. 
 The most extensive quenched studies have been carried 
out using either the non-relativistic (NRQCD) formulation \cite{cornell}
 or the heavy clover approach \cite{heavy} to $b$-quarks. 
 Results for $f_B$ and $f_{B_s}$ from these two 
methods are in good agreement with each other.

\vspace{.2in}
\noindent
In this article we investigate once again the NRQCD approach to 
leptonic decays of heavy mesons.  We wish to gain further insight into  
this method, which will strengthen confidence in our previous results
 and also allow us to assess its potential 
 when going on to other important areas 
of heavy quark physics such as semi-leptonic decays. 
Our analysis is based on quenched results for the $B_s$ meson 
decay constant from simulations at three lattice 
spacings spanning \ainvsp = 1.1GeV - 2.6GeV. We use NRQCD $b$-quarks,
 clover light quarks and the Wilson plaquette gauge action.
 Simulation parameters are summarized in 
Table~\ref{simdet}. 
The $\beta = 6.0$ $f_{B_s}$ results in sections III \& IV are taken from 
reference \cite{fbplb}.  The configurations at $\beta = 5.7$ and 
$\beta = 6.2$ plus the light propagators at $\beta = 6.2$ were 
provided by the UKQCD collaboration.  The $\beta=6.0$ data presented in 
section VI also uses UKQCD configurations and light propagators.
 Some preliminary results at earlier 
stages of this project have appeared in \cite{jhein98,sara99}.

\vspace{.1in}
\noindent
Our first goal was to check for independence of our final $f_{B_s}$ value
from the lattice spacing used.  This is particularly important here, since 
we employ an effective theory, NRQCD, for heavy quarks which 
precludes extrapolations to zero lattice spacing, $a \rightarrow 0$.  
There are many advantages 
to NRQCD, including being able to simulate quarks with masses much 
larger than the cutoff ($aM_Q \gg 1$) without introducing 
large discretization errors and the efficiency with which heavy 
quark propagators can be computed numerically.  However,
 the theory is not renormalizable and furthermore
 $aM_Q$ cannot become too small.  In order for the approach 
to work, one must have a window in which the lattice spacing is large enough 
so that $aM_Q $ is of $O(1)$ or greater, 
but  also small enough so that all discretization 
errors, including those coming from the glue and light quark sectors, are 
under control.  Physical results must already be 
independent of the lattice spacing within acceptable and quantifiable 
systematic errors.
 The scaling studies presented in this article demonstrate 
that such a window does indeed exist between \ainvsp = $1.1 \sim 2.6$ GeV for 
simulations with NRQCD-clover quarks and Wilson glue. The systematic 
errors are $\sim 10$\% in the upper half of the window and increase to
$ \sim 20$\% towards the coarse end.
 By further improving the quark and gauge actions one should be able to 
improve the errors at the coarse end and even 
 extend the window 
down below 1GeV.  Recently the JLQCD/Hiroshima collaboration 
\cite{jlqcdhiro} studied scaling of 
$f_B$ and $f_{B_s}$ 
with a lattice action very similar to the one used in this article. 
 Our results are in good agreement with theirs.

\vspace{.1in}
\noindent
Another feature of the use of an effective theory, is that the matrix 
element of current operators in the full theory (continuum QCD) such 
as the LHS of eq.(\ref{deffb}) must be built up out of several 
current matrix elements evaluated in the effective theory,
\be \label{amu}
\langle\,  A_\mu \, \rangle_{QCD} = \sum_j C_j \;
\langle \, J_\mu^{(j)} \, \rangle_{eff} .
\ee
With NRQCD heavy quarks the currents are of the form,
\be \label{jeffdef}
J_\mu^{(j)} = { 1 \over {(M_Q)^{l_j}}} O_\mu^{(j)} ,
\ee
where the $O^{(j)}_\mu$ are operators of dimension $d_j = 3 + l_j \;(l_j \ge
0)$. 
Quantum
corrections to the currents and the mixing of higher dimension 
operators back onto lower dimension operators will induce ``power law''
terms, i.e. terms that go as $\alpha_s/(aM_Q)^{\delta l_j}$
 (plus higher orders). 
  Since the LHS of eq.(\ref{amu}) should have no power law
contributions, any consistent framework for evaluating 
$\langle \, J_\mu^{(j)} \, \rangle_{eff}$ and $C_j$ must arrange for 
power law terms to cancel to the order one is working in.
The second goal of this article was to investigate and 
quantify this cancellation at $O(\alpha_s/(aM_Q))$.  In our simulations we 
include all relevant current operators through dimension 4. The matching 
coefficients $C_j$ have been calculated perturbatively through 
$O(\alpha_s/M_Q)$ and $O(a \, \alpha_s)$ \cite{pert1}. 
 Matrix elements of the dimension 4
current operators will contain in addition to $\Lambda_{QCD}/M_Q$ physical 
contributions also power law terms of which the dominant 
$O(\alpha_s/ (a \, M_Q))$ terms
 are cancelled by our matching coefficients.
Remaining uncancelled power law terms start at $O(\alpha_s^2/( a \, M_Q))$ 
 and their effects are taken into account
 in the systematic errors that we quote.
  Our analysis, based on the temporal component of 
the axial vector current, $A_0$, demonstrates that, although 
power law contributions are substantial there are no problems with cancelling 
them and there are no delicate fine tunings involved. 
Furthermore, we find that after cancellation of $O(\alpha_s/(aM_Q))$ 
terms the contributions from $1/M_Q$ current corrections are significantly 
reduced.

\vspace{.1in}
\noindent
The third goal of the present study was to carry out other consistency tests 
 in our lattice evaluation of \fbs.  
The definition  eq.(\ref{deffb}) of the decay constant 
tells us that one should be able to obtain the same \fbs from 
$B_s$ mesons at rest and from those with nonzero spatial momentum up to
 lattice artifacts.  We have simulated leptonic decays of $B_s$ mesons 
with nonzero momenta to check this.  Another project, which we will 
report on in a separate publication, is determining 
\fbs from the spatial component $A_k$, providing 
a further covariance test of calculations with NRQCD heavy quarks
\cite{fbfromak}.

\vspace{.1in}
\noindent
In the next section we introduce the lattice actions and current operators 
used in our simulations.  Section III presents results for $f_{B_s}$
 at the three 
lattice spacings and scaling plots.  In section IV we consider cancellation 
of power law contributions and also discuss 
 the slope of $f_{PS} \sqrt{M_{PS}}$ versus $1/M_Q$ as one 
leaves the static limit. 
 On our coarsest lattice we are able to simulate NRQCD 
charm quarks and in section V we present results for the $D_s$ meson decay  
constant $f_{D_s}$ at this one value for the lattice spacing. 
We describe 
 $f_{B_s}$ extracted at nonzero momenta 
in section VI and  section VII gives a brief summary of this article.
In an Appendix we provide a Table of heavy-light current 
matching coefficients covering a wider range of heavy quark masses 
than in \cite{pert1}.

\section{The Lattice Action and Current Operators }

\vspace{.2in}
\noindent
We used
	the NRQCD heavy quark action density given by, 
 \be \label{nrqcdact}
 \lag =  \overline{\psi}_t \psi_t - 
 \overline{\psi}_t
\left(1 \!-\!\frac{a \delta H}{2}\right)_t
 \left(1\!-\!\frac{aH_0}{2n}\right)^{n}_t
 U^\dagger_4
 \left(1\!-\!\frac{aH_0}{2n}\right)^{n}_{t-1}
\left(1\!-\!\frac{a\delta H}{2}\right)_{t-1} \psi_{t-1},
 \ee
where $\psi$ denotes a two-component Pauli spinor, 
 $H_0$ is the nonrelativistic kinetic energy operator,
 \be
 H_0 = - {\delsq\over2\Mbz},
 \ee
and $\delta H$ includes relativistic and finite-lattice-spacing
corrections,
 \begin{eqnarray}
\delta H 
&=& - c_1\,\frac{g}{2\Mbz}\,\sigmav\cdot\Bv \nl
& & + c_2\,\frac{ig}{8(\Mbz)^2}\left(\delv\cdot\Ev - \Ev\cdot\delv\right)
 - c_3\,\frac{g}{8(\Mbz)^2} \sigmav\cdot(\delv\times\Ev - \Ev\times\delv)\nl
& & - c_4\,\frac{(\delsq)^2}{8(\Mbz)^3} 
  + c_5\,\frac{a^2\delfour}{24\Mbz}  - c_6\,\frac{a(\delsq)^2}{16n(\Mbz)^2} .
\label{deltaH}
\end{eqnarray}
 In addition to all $1/M^2$ terms we have also included the leading 
$1/M^3$ relativistic correction 
as well as the discretization corrections appearing at the same 
order in the momentum expansion. 
For all three $\beta$ values 
 the NRQCD action was tadpole-improved using the 
plaquette definition of $u_0$ \cite{lepmac} and the $c_i$ were set equal 
to unity.  Table II lists the bare heavy quark masses $a M_0$ at which 
simulations were carried out together with 
the corresponding values for the stability parameter $n$.

\vspace{.1in}
\noindent
For the light quarks we used the tree-level tadpole improved clover action
 ($c_{SW} = u_0^{-3}$) at $\beta = 5.7$ and $\beta = 6.0$.  However at
 $\beta = 6.2$ 
the nonperturbatively determined value  $c_{SW}=1.61$ was employed 
\cite{alpha}.  The $\kappa$ values used in the simulations 
are listed in Table I.  At $\beta=5.7$ and $\beta=6.2$ we employed a single 
$\kappa$ close to $\kappa_s$ fixed through the $K$ meson.
 At $\beta=6.0$ data was obtained at 
three light $\kappa$'s, 0.1369, 0.1375 and 0.13808, and results were then 
interpolated to $\kappa_s$ \cite{fbplb}.
In Table I. we also list the lattice spacing determined from the $\rho$ mass 
and the bare dimensionless heavy quark mass $aM_0^b$ corresponding to the $b$
 quark. 
The latter was determined from a combination of perturbation 
theory for mass renormalization and simulation results for binding energies.
  These procedures for fixing 
$aM_0^b$ have been explained in several previous publications 
\cite{fbplb,joachim}.  The simulation parameters given in this paragraph 
apply to results presented 
in sections III, IV and V.  The $\beta=6.0$ simulations described in 
section VI use different parameters (see section VI and Table VII).

\vspace{.2in}
\noindent
For the temporal component of the heavy-light axial vector current 
eq.(\ref{amu}) becomes,
\be \label{a0lat}
{\langle  \, A_0 \, \rangle}_{QCD} = 
 \sum_{j=0}^2 C_j^{(A_0)} \langle \, J_0^{(j)} \, \rangle \;\; + \;\; O(1/M_Q^2) ,
\ee
with
\begin{eqnarray} \label{j0op}
 J^{(0)}_0 & = & \bar q \,\gamma_5 \gamma_0\, Q,\nonumber \\
 J^{(1)}_0 & = & \frac{-1}{2M_0} \bar q 
    \,\gamma_5 \gamma_0\,
\mbox{\boldmath$\gamma\!\cdot\!\nabla$} \, Q ,\nonumber \\
 J^{(2)}_0  & = & \frac{1}{2M_0} \bar q 
    \,\mbox{\boldmath$\gamma\!\cdot\!\overleftarrow{\nabla}$}
    \, \gamma_5 \gamma_0\, Q . \nonumber  \\
\end{eqnarray}
The heavy quark 4-spinor $Q$ has the 
NRQCD 2-spinor $\psi$ as 
the upper two components, zero for the lower two components and obeys 
$\gamma_0 Q = Q$. 
$J_0^{(0)}$ and $J_0^{(1)}$ contribute already at  tree-level
whereas $J_0^{(2)}$ appears only at one-loop. 

\vspace{.1in}
\noindent
Higher dimension operators appear in lattice
 effective theories when one, for instance, needs to consider 
relativistic corrections.  They also occur whenever lattice 
operators are improved. Improvements to current operators
 take on the form,
\be \label{jimpdef}
J_\mu^{(j)disc} = a^{l_j} O_\mu^{(j)} ,
\ee
where again, as in eq.(\ref{jeffdef}),
 the $O^{(j)}_\mu$ are operators of dimension $d_j = 3 + l_j$. In the 
present case of NRQCD-clover currents, one has the identical 
set of possible dimension 4 operators $O_\mu^{(j)}$ in eq.(\ref{jeffdef}) and 
eq.(\ref{jimpdef}).  Hence a one-loop mixing and matching calculation to 
full continuum QCD involving the above $J_\mu^{(j)}$  will 
not only tell us how the $1/M_Q$ currents contribute  but will 
also identify the discretization corrections that come in at one-loop. 
 In other words a 
consistent matching through $O(\alpha_s/M_Q)$ should automatically be 
consistent through $O(a\,\alpha_s)$.  In reference \cite{pert1} 
it was found that for each $\mu$ only
 one of the $J_\mu^{(j)disc}$ 
was relevant at one-loop.  Matching to continuum QCD 
could be achieved by 
improving $J^{(0)}_\mu$ in the following way,
\be 
\label{jimp}
J_\mu^{(0)} \quad  \rightarrow \quad J_\mu^{(0)} + 
\alpha_s \, \zeta_{A_\mu} \,J_\mu^{(disc)} ,
\ee
with 
\be
\label{jdisc}
 J^{(disc)}_\mu   =  - a\, \bar q 
    \,\mbox{\boldmath$\gamma\!\cdot\!\overleftarrow{\nabla}$}
    \, \gamma_0 \gamma_5 \gamma_\mu\, Q  = 2aM_0\,J_\mu^{(2)} ,
\ee
and $\zeta_{A_0} = 1.03$.
 The effect of $J_\mu^{(disc)} $ can be taken into account by 
expressing it in 
terms of $J_\mu^{(2)}$ and absorbing  $\alpha_s 2aM_0\,\zeta_{A_\mu}$ into 
the coefficient $C_2^{(A_\mu)}$. 

\vspace{.1in}
\noindent
At $\beta = 5.7$ and $\beta = 6.0$ $1/M_Q^2$ current corrections that appear 
at tree-level have also been studied. For $A_0$ these are,
\begin{eqnarray} \label{jtree}
 J^{(3)}_0 & = & 
 \frac{1}{8 M_0^2}\,\bar{q}\,\gamma_5\gamma_0 \delvc^2 \, Q  ,\nonumber \\
 J^{(4)}_0 & = & 
 \frac{g}{8 M_0^2}\,\bar{q}\, \gamma_5 \gamma_0  \Sigmav \cdot \Bv 
\, Q , \nonumber \\
 J^{(5)}_0 & = & \frac{-ig}{4 M_0^2} \,\bar{q}\, \gamma_5 \gamma_0
 \alphav \cdot \Ev  \,Q,
\end{eqnarray}
where $\alphav \equiv \gamma_0 \gammav$ and  $\Sigmav = 
diag(\sigmav, \sigmav)$.  
We  will comment on matrix elements of these currents and their $M_Q$ 
dependence later in section V.  However, since
the $O(\alpha_s/M_Q^2)$ full matching calculation has not been 
carried out yet, $O(\alpha_s/(a\,M_Q)^2)$ and $O(\alpha_s 
(\Lambda_{QCD}/M_Q)/(a \, M_Q))$ 
power law contaminations contained in these matrix elements 
 will not be cancelled and  we prefer not to 
include them in our scaling analysis of \fbs.  
  In reference \cite{fbplb} 
tree-level contributions from (\ref{jtree}) were included.  The 
total effect was a $\sim 3$\% decrease in \fbs, well within the 
systematic errors quoted there for uncancelled power law terms.  
In the $\beta=6.0$ results given in the present article, however, 
we have removed the $1/M_Q^2$ tree-level current effects.

\vspace{.1in}
\noindent
Based on  the lattice actions employed, the improvement of $J^{(0)}_\mu$ 
and the 
consistent matching to continuum full QCD through $O(\alpha_s/M_Q)$ and 
$O(a \, \alpha_s)$ in the currents, one can now list 
 the expected remaining dominant systematic errors 
in our evaluation of $\langle \,  A_\mu \, \rangle_{QCD}$.  These are 
collected in Table III and their size, in percent, estimated for each $\beta$. 
In making these estimates we use $\Lambda_{QCD} \sim 400MeV$, $\alpha_s(\beta
=5.7) \sim 0.33$, $\alpha_s(\beta=6.0) \sim 0.24$, $\alpha_s(\beta = 6.2) 
\sim 0.20$ and approximate $aM_Q$ with $aM_0^b$ of Table I.  The 
$\alpha_s$ values are close to $\alpha_P(q^\ast = 1/a)$, where $\alpha_P$ 
is the coupling introduced in reference \cite{nrqcdalpha} 
 closely related 
to $\alpha_V$ which is based on the heavy quark potential \cite{blm,lepmac}.
Discretization errors come in at $O((a \Lambda_{QCD})^2)$ from the currents 
and at $\beta = 5.7$ and $\beta=6.0$ also at $O(\alpha_s \, a \Lambda_{QCD})$ 
from the light quark action. 
These latter errors arise from using a tadpole-improved $c_{SW}$ rather 
than a non-perturbative $c_{SW}$ (or a $c_{SW}$ value fully corrected 
at one-loop) and are therefore proportional 
to the difference of these values for $c_{SW}$. Such differences and, hence, 
also the $O(\alpha_s \, a \Lambda_{QCD})$ errors are small at $\beta = 6.0$.
Relativistic 
errors are at $O((\Lambda_{QCD}/M_Q)^2)$ from the currents.
 There are also $O(\alpha_s\,\Lambda_{QCD}/M_Q)$ errors from using 
the tadpole-improved coefficient for the $\sigma \cdot B$ term in the NRQCD
action instead of the full $O(\alpha_s)$ coefficient matched 
to continuum QCD. This error is proportional to the $1/M_Q$ slope 
induced in $f_B$ by the $\sigma \cdot B$ term. The investigations in 
\cite{sara} found this slope to be small, 
approximately $0.25*\Lambda_{QCD}/M_Q$.
We fold in perturbative errors 
separately for 
 $O(\alpha_s^2)$ 
 and $O(\alpha_s^2/(aM_Q))$ since these have different $\beta$ dependence 
and also to indicate that perturbation theory will eventually break down 
for small $aM_Q$, i.e. for large enough $\beta$.
We note that in the absence of  tree-level $O(1/(M_Q^2)$ current 
matrix elements, 
 there are no $O(\alpha_s/(aM_Q)^2)$ errors in our calculations.  As long 
as the same action is used in both the simulations and the perturbative 
matchings, all $O(\alpha_s/(aM_Q)^j)$ terms  cancel between matrix 
elements and matching coefficients.
Table III also lists other systematic errors (uncertainty in $\kappa_s$ and 
fixing of \ainvsp from the $\rho$ mass) 
and the statistical errors 
in the \fbs determinations.  Not listed are uncertainties due to quenching. 
Recent studies indicate that these last corrections can be substantial, at the 
$10 \sim 25$\% level \cite{sara,cppacs1,cppacs2}.

\section{Simulation Results and Scaling of \fbs }

In this section we describe \fbs calculated from $A_0$ and investigate 
scaling of this quantity.  
 We start with simulation results for current matrix elements 
in lattice units before renormalization.  
Table~\ref{treefj}  summarizes our results at 
$\beta = 5.7$ and $\beta=6.2$ for 
\be 
\label{fsqrtmdef}
a^{3/2}\, f^{(j)}\,\sqrt{M_{PS}} = \frac{a^{3/2}}{\sqrt{M_{PS}}} \, 
 \langle 0|\, J_0^{(j)} \,|PS \rangle .
\ee
Data are presented for several heavy quark masses with the light 
quark mass fixed to $\kappa_q$ of Table I.  The corresponding 
simulation data for $\beta = 6.0$ can be found in \cite{fbplb}.  
The matrix elements were evaluated in the 
meson rest frame.  $\langle \,J_0^{(2)} \, \rangle$ is identical to 
$\langle \, J_0^{(1)} \, \rangle$ in this frame 
and hence is not listed separately in the Table.
For $\beta=5.7$ we also list results for the higher order currents of 
eq.(\ref{jtree}).

\vspace{.2in}
\noindent
Results in physical units ($GeV^{3/2}$) for 
\be
 f_{PS}\,\sqrt{M_{PS}} = \sum_{j=0}^2 C_j^{(A_0)} f^{(j)}\, \sqrt{M_{PS}}
\ee
 are summarized in Table V, at tree-level ($C_0^{(A_0)} = C_1^{(A_0)} = 1$, 
$C_2^{(A_0)} = 0$) and with one-loop matching for three
 different scales to fix $\alpha_s = 
\alpha_P(q^\ast)$.  At $\beta=5.7$ and $\beta=6.0$ we can directly take 
over the matching coefficients from Appendix A, which 
 assume tadpole improvement in both the NRQCD and clover 
actions.  At $\beta = 6.2$ where the non-perturbative $c_{SW}$ is
used, some slight modifications are necessary.  We first remove the 
tadpole improvement term coming from the light quark wave function 
renormalization (denoted $C_q^{TI}$ in \cite{pert1}) from the matching 
coefficients. We then use $\kappa$ rather than 
$\tilde{\kappa} = \kappa \times u_0$  to rescale the light quark 
field in the currents.  As a result the entries in Table IV and the
tree-level results in Table V 
are enhanced for $\beta=6.2$ by approximately a factor of $u_0^{-1/2}$.
 This enhancement is 
 compensated for to a large extent upon going to renormalized matrix elements 
(the latter three columns in Table V) due 
to larger perturbative matching coefficients. Alternatively, we could have 
used another light quark 
action at $\beta=6.2$ which has the first derivative terms 
divided by $u_0$ and a clover term with  coefficient
 $C_{SW}^{nonpert.}/u_0$.  This would have allowed us to use tadpole-improved 
perturbation theory even for the light quark propagator, similar to what was
done at the other 
$\beta$ values.  We tested this option and found that the final answer 
for the decay constant changed by $4 \sim 5$\%.  This difference is,
 as expected,  of the 
same order as other $O(\alpha_s^2)$  corrections not included in the present 
analysis.

\vspace{.1in}
\noindent
The scale $q^\ast$ has not been calculated yet for the matching coefficients 
with NRQCD heavy and clover light quarks. 
$q^\ast$ is known for static heavy quarks combined with 
massless Wilson light quarks 
 \cite{herhill} and also for 
 light-light currents \cite{craige} again using massless 
Wilson fermions.  
These authors find $q^\ast = 2.18/a$ and $q^\ast = 2.3/a$ respectively, 
suggesting very mild dependence of $q^\ast$ on $aM_Q$.  Results 
also exist for clover light-light currents \cite{crisa} showing 
some reduction in $q^\ast$ relative to the Wilson results for 
these currents.  
In the absence of an explicit calculation for the action and currents
 employed in the present simulations, 
we use the $q^\ast = 2/a$ numbers to quote central values and arrange 
for perturbative errors to cover the spread in results due 
to uncertainty in $q^\ast$.  
Table VI shows our final numbers for $f_{B_s}$ interpolated to 
the physical heavy quark mass $M_0^b$ for the three lattice spacings. 
The Particle Data Group's $B^0_s$ mass, $5.369$GeV, was used to convert from 
$f_{PS} \sqrt{M_{PS}}$ to $f_{B_s}$.  The errors follow those listed in 
Table III and correspond from left to right to statistical (plus interpolation
in $M_0$), discretization, 
perturbative, relativistic, $\kappa_s$ determination and errors in
\ainv($m_\rho$). 

\vspace{.1in}
\noindent
After adding all errors in quadrature,
 we plot 
 $f_{B_s}$ versus $a$ in Fig. 1  and 
 compare with results from \cite{jlqcdhiro}.
One finds good scaling within errors and excellent agreement between the two 
collaborations.  These results also agree well with quenched 
$f_{B_s}$ determinations 
using other heavy fermion formulations on the lattice \cite{fb}.  

\vspace{.1in}
\noindent
Table VI and Fig.1 represent one of the main results of this article. 
Working around \ainvsp of $\sim 2$GeV with NRQCD-clover quarks and 
Wilson glue and matching through $O(\alpha_s/M_Q)$ and 
$O(a \, \alpha_s)$, one can obtain reliable quenched numbers for 
B meson decay constants with $\sim 10$\% accuracy.  To do better at 
these lattice spacings one would need to go to higher orders in the 
matching procedure.  At our coarsest lattice spacing the systematic 
uncertainties are larger. Going to more highly  improved 
actions would help here.
Nevertheless, the $\beta = 5.7$ 
results are consistent with our best numbers.  Simulations with 
\ainvsp$\sim 1$GeV can already give a good indication of continuum 
physics.

\vspace{.1in}
\noindent
Since we cannot extrapolate to $a \rightarrow 0$,
in order to quote our best estimate for quenched \fbs there are several 
choices.  One approach would be to
 average over the $\beta=6.0$ and $\beta=6.2$ results.  Alternatively one 
could, after having verified scaling behaviour, 
 pick the point with the smallest statistical and systematic errors.  Based on 
Table III this corresponds to the $\beta=6.2$ result.  We adopt the second 
method here and present as our final best value for the $B_s$ meson 
decay constant in the quenched approximation,
\be\label{best}
f_{B_s} = 187(4)(4)(11)(2)(7)(6) \; {\rm MeV} ,
\ee
with errors as described in Table VI.

\vspace{.1in}
\noindent
If the value of $f_{B_s}$ at the physical $b-$quark mass were our only goal 
then we would be done at this point. 
 However, if further details are of interest,
 such as 
`` the true size of contributions from $1/M_Q$ current 
corrections to 
$f_B$ ''  or ``  the slope of $f_{PS}\sqrt{M_{PS}}$ versus 
$1/M_Q$ '', then one needs to scrutinize more carefully the 
 contributions from individual 
current matrix elements $\langle \, J_0^{(j)} \, \rangle$ 
and matching coefficients $C_j$ to the RHS of eq.(\ref{a0lat}).
We turn to these issues in the next section.

\section{Power Law Terms and $\Lambda_{QCD} / M_Q$ Corrections}
In this section we investigate further the properties of matrix 
elements of the $1/M_Q$ current corrections.  We focus 
on the current 
$ J^{(1)}_0  =  \frac{-1}{2M_0} \bar q 
    \,\gamma_5 \gamma_0\,
\mbox{\boldmath$\gamma\!\cdot\!\nabla$} \, Q $.
This operator is introduced in order to take into account 
$p/M_Q$ effects which we know are present in the full theory, i.e. 
in the LHS of eq.(\ref{a0lat}).  Here ``$p$'' is the momentum of the 
quarks inside the meson and should be of $O(\Lambda_{QCD})$.  Once 
introduced into the effective theory, however, this current will mix
with other currents in the theory.  Its short distance effect will 
be to renormalize the zeroth order current and one expects the 
matrix element $\langle \, J^{(1)}_0 \,\rangle$ to develop a 
component proportional to 
$\langle \, J^{(0)}_0 \, \rangle$. 
 On dimensional grounds 
the proportionality coefficient will be of the form $const/(a\, M_Q)$ 
and one has,
\be \label{powerlaw}
\langle \, J^{(1)}_0 \, \rangle =  \frac{C_{10}}{a \, M_Q} 
\langle \, J_0^{(0)} \, \rangle + \frac{\Lambda_{QCD}}{M_Q} \; term 
\qquad + \qquad higher \; orders.
\ee

\vspace{.1in}
\noindent
The first term on the RHS of (\ref{powerlaw})
is an example of a power law contribution 
to a matrix element.  They are unavoidable in a quantum effective field
 theory where one has mixing between operators of different dimensions.  
However, the matching procedures between the full 
and the effective theories should remove 
these artifacts.  Since we do the matching perturbatively, power law 
contributions can only be subtracted to a given order in $\alpha_s$.  
As mentioned several times already, in our present calculations we 
cancel terms through $O(\alpha_s/(a\,M_Q))$.
To see precisely how this cancellation comes about we need to dissect the 
matching coefficient $C_{j=0}$.  
Using the notation of \cite{pert1} one can rewrite eq.(\ref{a0lat}) as, 
\be \label{ajlat}
{\langle  \, A_0 \, \rangle}_{QCD}
=(1 + \alpha_s \, \rho_0) \, \langle J_0^{(0)}\rangle 
+(1 + \alpha_s \, \rho_1) \,\langle J_0^{(1)}\rangle 
+ \alpha_s \, \rho_2 \;\, \langle J_0^{(2)}\rangle ,
\ee
with 
\be \label{rho0}
\rho_0  =  B_0 
 - \frac{1}{2}(C_q+C_Q)
 - \zeta_{00} - \zeta_{10} \equiv \tilde{\rho}_0 - \zeta_{10}.
\ee
$\rho_0$ is the one-loop contribution to $C_0$ and consists of parts
coming from continuum perturbation theory ($B_0$), from light and heavy quark 
wave function renormalizations on the lattice ($C_q$ and $C_Q$) and 
from mixing between the currents $J_0^{(j)}$ ($\zeta_{00}$ and 
$\zeta_{10}$). $\zeta_{00}$ is the feedback of $J^{(0)}_0$ onto itself 
and $\zeta_{10}$ incapsulates the projection of $J^{(1)}_0$ back onto 
$J^{(0)}_0$, precisely what one is looking for in the first term of 
(\ref{powerlaw}).  
Picking out the relevant terms, one can write 
\begin{eqnarray} \label{powersub}
(1 + \alpha_s \, \rho_0) \, \langle J_0^{(0)}\rangle 
 + \langle J_0^{(1)}\rangle  &=& 
(1 + \alpha_s \, \tilde{\rho}_0) \, \langle J_0^{(0)}\rangle \nl
& & + \left[ \frac{C_{10}}{a \, M_Q} - \alpha_s \, \zeta_{10} \right] 
\langle \, J^{(0)}_0 \, \rangle
 + \frac{\Lambda_{QCD}}{M_Q} \; term \; + \; \ldots
\end{eqnarray}
The power law contribution 
$ \left[ \frac{C_{10}}{a \, M_Q} - \alpha_s \, \zeta_{10} \right] 
\langle \, J^{(0)}_0 \, \rangle $ has now been reduced from 
an $O(\alpha_s/(a\,M_Q))$ to an $O(\alpha_s^2/(a \, M_Q))$ effect through 
the $\alpha_s \zeta_{10}$ matching term. 
It is convenient to define a subtracted $J_0^{(1)}$, 
\be \label{jsub}
\langle\,J_0^{(1)} \,\rangle_{sub} = 
\langle\,J_0^{(1)} \,\rangle  - \alpha_s \, \zeta_{10} \,
\langle\,J_0^{(0)} \,\rangle,
\ee
which is equal to the last two terms in (\ref{powersub}) up to higher 
order corrections.  It is 
$\langle\,J_0^{(1)} \,\rangle_{sub}$ rather than 
$\langle\,J_0^{(1)} \,\rangle$ that should be used 
to assess the importance  of $1/M_Q$ current corrections to $f_{B_s}$. 
In Fig.2 we plot the absolute values of 
$f^{(0)}\sqrt{M}$, 
$f^{(1)}\sqrt{M}$ and 
$f^{(1)}_{sub}\sqrt{M}$ 
(related to $\langle\,J_0^{(0)} \,\rangle$,
$\langle\,J_0^{(1)} \,\rangle$ and 
$\langle\,J_0^{(1)} \,\rangle_{sub}$ in the usual way)  versus $1/M_{PS}$
at $\beta=6.0$.  One sees that 
$|f^{(1)}_{sub}|\sqrt{M}$ is considerably smaller than 
$|f^{(1)}|\sqrt{M}$ and amounts to only $\sim4$\% to $\sim5$\% of the 
lowest order $f^{(0)}\sqrt{M}$ term.  This is of the order of 
 systematic errors such as $O(\alpha_s^2)$ errors remaining 
in our calculation, so we cannot give a more precise estimate of the 
true $\Lambda_{QCD}/M_Q$ contributions coming from $1/M_Q$ currents.
Similar results are obtained at 
the other lattice spacings.  A corresponding plot similar 
to Fig.2 for $\beta=5.7$
will be discussed in the next section.
At all three $\beta$ values the power law 
contributions make up 50\% to 80\% of 
$\langle\,J_0^{(1)} \,\rangle$, depending on $q^\ast$,
 and once they are subtracted away one 
is left with only a small remainder.
We note that 
 the subtraction in (\ref{jsub}) took place automatically in the 
$f_{B_s}$ calculations of the previous section.  The fact that 
$f^{(1)}_{sub}\sqrt{M}$ turned out to be small did not pose any 
problems for 
the power law cancellation itself at $O(\alpha_s/(aM_Q))$.

\vspace{.1in}
\noindent
Another quantity where having information on $\Lambda_{QCD}/M_Q$ 
corrections is important, is the slope of $f_{PS}\sqrt{M_{PS}}$ versus
 $1/M_Q$.  Writing,
\be \label{hqetexp}
f_{PS}\sqrt{M_{PS}} = 
\left(f_{PS}\sqrt{M_{PS}}\right)_{stat} \, [ 1 + 
\frac{\Lambda_{sl}}{M_Q} + O(1/M_Q^2) ] ,
\ee 
one is interested in the slope $\Lambda_{sl}$. 
 The bulk of the effect will 
come from $1/M_Q$ terms in the NRQCD action which are responsible for 
the $M_Q$ dependence of the zeroth order current matrix element 
 $\langle\,J_0^{(0)} \,\rangle $, and from the 
one-loop correction $\tilde{\rho}_0$ to this matrix element. 
 We have just seen that contributions 
from $1/M_Q$ current corrections are small.  Nevertheless, 
when simulation data are fit to the form (\ref{hqetexp}) to extract 
$\Lambda_{sl}$ one might worry about contaminations
 due to uncancelled power law terms 
of $O(\alpha^2_s/(a\,M_Q))$.  
If $\Lambda_{sl}$ is the result of a fit then one can estimate a 
percentage error of $\alpha_s^2/(a \, \Lambda_{sl})$ in 
this slope. Putting in some numbers this leads to 
$\frac{10 \sim 13}{\Lambda_{sl}[GeV]}$\% uncertainties
depending on $\beta$,
 where $\Lambda_{sl}$ must be inserted in GeV's.  
If the slope is less than $\Lambda_{QCD}$ ($ 0.4$ GeV) these errors can
become quite large.

\vspace{.1in}
\noindent
There is, however, another reason why it is difficult to extract an 
accurate $\Lambda_{sl}$ from our simulations. In order to calculate 
the slope we need to look at data at masses larger than the $b$ quark 
mass. The one-loop terms in the matching coefficients increase as 
one goes towards the static limit and as a consequence results for 
$f_{PS}\sqrt{M_{PS}}$ become much more sensitive to the value of $\alpha_s$ 
i.e. to $q^\ast$.  This can be seen in Fig. 3 where 
$f_{PS}\sqrt{M_{PS}}$ is plotted versus $1/M_{PS}$ for $q^\ast = 1/a$ and 
$q^\ast = \pi/a$.  One sees that there is considerable dependence of the 
 slope on  $q^\ast$.  We will quote a central 
value for the slope based on the $\beta=6.0$ 
data using $q^\ast = 2/a$ and allow for a change in $q^\ast$ in our error. 
We note that in order for the expansion in (\ref{hqetexp}) to be 
 valid  there should be  only power dependence and 
 no logarithmic dependence on $M_Q$.  We satisfy this condition by 
fixing the logarithms in the matching coefficients to log$(a\, M_0^b)$.  
Our best estimate for $\Lambda_{sl}$ is then:
$\Lambda_{sl} = -0.67(9)(^{-17}_{+34})(12)$GeV.  The first error is 
the statistical fit error to the $q^\ast = 2/a$ data and the second comes from 
the upper limit of errors on the magnitude of the slope 
 from fits to 
$q^\ast=\pi/a$ data and the lower limit of errors in fits to $q^\ast=1/a$ data.
 The third error is an estimate 
for $\alpha_s^2$ and $\alpha_s^2/(a\,M_Q)$ corrections.
In Fig. 4 we plot the $\beta = 6.0$ results together with data from 
other $\beta$ values all using $q^\ast=2/a$. 
 One sees that slopes extracted at the two finer 
 lattice spacings are consistent within our large errors.  Results from 
 $\beta=5.7$ 
 lie mostly parallel to the $\beta=6.0$ and $\beta=6.2$ 
 data, however 
 statistical errors are too large near the static limit to enable a useful
 estimate for the slope.  Around the physical B meson the $\beta=5.7$ data 
points lie $\sim$15\% high, in agreement with results in Table VI and 
consistent with our estimates of systematic errors in Table III.

\vspace{.1in}
\noindent
In contrast to $f_{B_s}$ which we demonstrated can be calculated with 
$\sim 10$\% errors, we find that specific $1/M_Q$ corrections are 
harder to determine with comparable percentage accuracy. 
  For the two quantities discussed in this 
section, the $1/M_Q$ corrections coming from the current corrections and 
the slope of $f_{PS}\sqrt{M_{PS}}$ as one leaves the static limit, the main 
reason is uncertainty coming from higher order perturbative corrections.

\section{Simulations at the Charm Quark and \fds}

In this section we discuss the behaviour of the decay constant and the
current matrix elements for heavy quark masses in the charm
region. For the coarsest of our lattice spacings the charm quark can
be reached at a bare mass of $aM_0^c = 0.87(6)(3)(^{+0}_{-13})$
\cite{joachim}.

\vspace{.1in}
\noindent
The size of the lattice current matrix elements for different values
of the heavy quark mass is compared in Fig.5 .
This figure is similar to Fig.2, however at $\beta=5.7$ we cover 
a much wider range of mass values.  We also plot $1/M_Q^2$ current 
matrix elements.
 The figure shows the value of
$|f^{(1)}_{sub}|\sqrt{M}$ to be subleading even in the charm region,
however it has grown from 4\% of $f^{(0)}\sqrt{M}$ at the $B_s$ to
14\% at the $D_s$. For $aM_Q \approx 1$ the value of the $1/M_Q^2$ current 
matrix element 
$(|f^{(3)}+f^{(4)}+f^{(5)}|)\sqrt{M}$ is equal to
$|f^{(1)}|\sqrt{M}$ and the two lines in Fig.5 cross. 
 After the discussion in the previous section, this is not
unexpected and provides evidence that this current is also dominated
 by power law contributions. The part of the current
$\vev{J_0^{(3)}+J_0^{(4)}+J_0^{(5)}}$ proportional to
$\vev{J_0^{(0)}}$ diverges as $\alpha_s/(aM_Q)^2$. For $aM_0 \approx
1$ this results in a suppression of ${\cal O}(\alpha_s)$ with respect
to $\vev{J_0^{(0)}}$, which is the same suppression factor as the one
for $\vev{J_0^{(1)}}$.  In
order to avoid uncertainties of ${\cal O}(\alpha_s/(aM_Q)^2)$ in the
final $f_{D_s}$, we do not include the matrix elements of $1/M_Q^2$ currents.
 By doing so, we have to accept
an uncertainty of the size of $\Lambda_{QCD}^2/M_Q^2$, which is 10\%,
for the final result.  This is a conservative estimate, however.
 Based on our experience with $\Lambda_{QCD}/M_Q$ current
 corrections, we can expect that most of 
 $(|f^{(3)}+f^{(4)}+f^{(5)}|)\sqrt{M}$ plotted in Fig.5 will eventually 
be cancelled 
by matching coefficients at $O(\alpha_s/(aM_Q)^2)$.  So, even at the 
$D_s$, the true 
 $\Lambda_{QCD}^2/M_Q^2$ current corrections should not be a large effect.  
The other 
 $\Lambda_{QCD}^2/M_Q^2$ contributions in $f_{D_s}$ come from the 
 $\Lambda_{QCD}^2/M_Q^2$ terms in the NRQCD action, and these are 
already included in the present simulations.  To estimate the total 
NRQCD error we add to the 10\% another 2\% coming from $O(\alpha_s \,
\Lambda_{QCD}/M_Q)$ errors in the NRQCD action (see discussion at the 
end of section II).

For the perturbative uncertainties we assign 11\% for
$\alpha^2_s$. This is larger than the 
uncertainty of $f_{D_s}$ arising from the variation of $q^*$ in a
range from $1$ to $\pi$. For the residual power law
contributions of ${\cal O}(\alpha_s^2/aM_Q)$ we assign an uncertainty of
 13\%, which leads to
17\% for the total perturbative error, 
when added in quadrature. The 13\% uncertainty 
is comparable to the size of the entire 
$\vev{J_0^{(1)}}_{sub}$ matrix 
element. For the uncertainty in the strange quark mass we assign 4\%
as in the case of $f_{B_s}$, which turns out to be negligible compared to the
other sources of uncertainty. It is interesting to note that the
uncertainty arising from the scale $a(m_\rho)$ almost cancels between
the effect this has on the conversion of the matrix element into
physical units and the shift of the bare charm quark mass.

\vspace{.1in}
\noindent
For our final result we obtain
\be \label{fdsresult}
f_{D_s} = 223(6)(31)(38)(23)(9)(^{+3}_{ -1})\; {\rm MeV}
\hspace{1cm}\rightarrow\hspace{1cm} f_{D_s} = 223(55)\;{\rm MeV}\,.
\ee
The uncertainties are listed in the same order as in Table~\ref{errors}
with the two discretization errors added in quadrature. This
result is in very good agreement with the lattice average of 
$220^{+25}_{-20}$ MeV quoted in \cite{draper}  for quenched $f_{D_s}$, 
 however the uncertainty we assign is larger.  It is also interesting 
to compare with direct experimental determinations of $f_{D_s}$. 
In a recent review 
article the authors of reference \cite{expfds} present a world averaged
experimental value of $f_{D_s} = (241 \pm 32)$MeV, which is very consistent 
with the lattice numbers.  Recall, however, that the lattice results given here
are all in the quenched approximation.

\section{\fbs from Finite Momentum Mesons }
This section  investigates the defining matrix element eq.(1) for nonzero 
spatial momentum
$\vec{p}$ at two lattice spacings, $\beta=6.0$ and $\beta=6.2$.  
Our goal 
is not to calculate  $f_{B_s}$ more accurately than we have done in 
section III from B mesons at rest,  but to test our ability to simulate 
hadrons at finite spatial momenta and obtain consistent results.  
At each $\beta$ we work at one 
value for the heavy quark mass and we do not attempt to extrapolate 
to the physical $b$ quark.  Furthermore at 
$\beta= 6.0$  we use a  different set of configurations from 
the previous sections.  The new configurations and light propagators 
were provided by the UKQCD collaboration.  In Table VII we summarise relevant 
action parameters.

\vspace{.1in}
\noindent
We accumulated data for four nonzero momenta, (0,0,1), (0,1,1), (1,1,1) 
and (0,0,2)
in units of $2 \pi/aL$ averaging over all equivalent momenta $\pm p_x$, 
$\pm p_y$ etc.   From the difference of falloff energies of meson 
correlations with and without momenta,
\be
\delta E(p) \equiv E_{sim}(p) - E_{sim}(0) ,
\ee
one can use a relativistic dispersion relation to define a mass for 
the heavy meson, usually called $M_{kin}$,
\be
M_{kin} = (p^2 - \delta E^2) / (2 \,\delta E)  .
\ee
Our first test is to check the extent to which $M_{kin}$ is independent 
of $p$. 
This nonperturbatively determined mass can also be compared with another mass
$M_{pert}$ 
based on the perturbative pole mass for the $b$ quark combined with an
energy shift $E_0$ and simulation results for $E_{sim}(p=0)$,
\be
M_{pert} = Z_m  M_0 - E_0 + E_{sim}(0) .
\ee
We use one-loop perturbation theory for $Z_m$ and $E_0$.   The one-loop 
coefficients and $q^\ast$ for the combination 
$(Z_m M_0 - E_0)$ are given in \cite{joachim}.
In Table VIII we list results for $M_{kin}$ and $M_{pert}$   in lattice 
units.  The errors in $M_{kin}$ are statistical and for 
$M_{pert}$ we give  $O(\alpha_s^2)$ perturbative errors.  The statistical 
errors in $M_{pert}$ coming from $E_{sim}(0)$ are negligible compared to 
the perturbative errors. The quality of our signals for finite momentum mesons 
depended rather strongly on the smearings employed 
(see \cite{joachim} for details of smearings).
 In some instances, e.g. in the 
$\beta=6.0$ data, clearer signals were obtained for the highest momentum than 
for some of the lower ones. 
We believe this is 
because the smearings used were better suited for that momentum.  For 
momentum (1,1,1), on the other hand, we could not extract any useful 
results at $\beta=6.0$.
 Fig. 6 plots $aM_{kin}$ versus the momentum and 
compares with $aM_{pert}$.  $M_{kin}$ is reasonably independent of 
$p$ for the range considered and agrees with $M_{pert}$ within 
$\sim$$1 \sigma$.  The statistical errors
at some of our data points, however, are still considerable for 
$M_{kin}$ and they could be hiding systematic discretization corrections.

\vspace{.1in}
\noindent
The next quantity of interest is the ratio of the decay matrix element
for a heavy meson with momentum $\vec{p}$ to that of a heavy meson 
decaying at rest \cite{simone}.  More specifically we consider,
\be\label{ratio}
 \frac{\langle 0 | A_0 | PS \;,\; \vec{p} \rangle /\sqrt{E(p)}}
{\langle 0 | A_0 | PS \;,\; \vec{p}=0 \rangle /\sqrt{M_{PS}}}
= \frac{\sqrt{E(p)}}{\sqrt{M_{PS}}} ,
\ee
where $E(p)$ is the total energy of the meson. 
The RHS of eq.(\ref{ratio}) is a very slowly rising function of 
$p$, never getting very much above 1 for the momenta involved. 
In our simulations we take the following expression for the LHS, working 
to the same order in $1/M_Q$ as in the previous sections.
\be\label{ratiof}
R(p) \equiv \frac{\sum_{j=0}^2 C^{(A_0)}_j\langle 0 | J_0^{(j)} |  
\vec{p} \rangle /\sqrt{E(p)}} 
{\sum_{j=0}^2 C^{(A_0)}_j
\langle 0 | J_0^{(j)} |  \vec{p}=0 \rangle /\sqrt{M_{PS}}} .
\ee
Whereas $M_{kin}$, discussed above, checks for the correct 
relativistic dispersion relation in the energies, $R(p)$ tests the $p$ 
dependence of amplitudes.
In Table IX we give  results for $R(p)$.  We also list $R^{(0)}(p)$, the 
analogous ratio using only the zeroth order current $J_0^{(0)}$,
\be\label{ratio0}
R^{(0)}(p) \equiv \frac{\langle 0 | J_0^{(0)} |  
\vec{p} \rangle /\sqrt{E(p)}} 
{\langle 0 | J_0^{(0)} |  \vec{p}=0 \rangle /\sqrt{M_{PS}}} .
\ee
One sees only a small change at 
less than the $1 \sigma$ level between $R(p)$ and 
$R^{(0)}(p)$.  Most of the effect of one-loop matching and $1/M_Q$ current 
corrections is canceled in this ratio.
In Fig.7 we plot $R(p)$ versus the momentum.  The full line is the 
expected behaviour coming from the RHS of eq.(\ref{ratio}).  A single line 
suffices for the two $\beta$ values,  since the difference in $M_{PS}$ 
is negligible.  One sees that results consistent with continuum expectations
 are obtained up to momenta of about 1.2 GeV.  
Even for the largest momenta
 deviations are  less than $\sim$8\%.

\section{Summary}
In this article we presented further tests of \fbs determinations using 
NRQCD heavy quarks and clover light quarks. We checked for scaling, 
studied the dependence on the momentum of the decaying $B_s$ meson and
also investigated cancellation of power law terms through $O(\alpha_s/(aM))$.
At one value of the lattice spacing we were able to determine the 
$D_s$ meson decay constant. Our best values for \fbs and 
 $f_{D_s}$ are given in eq.(\ref{best}) and eq.(\ref{fdsresult}) respectively.
Our results are in good agreement with other lattice determinations of 
these quantities. In the case of
 $f_{D_s}$, for which experimental measurements exist, 
we obtain a value consistent with the current experimental world 
average as given in \cite{expfds}.

\acknowledgements

This work was supported by the DOE under DE-FG02-91ER40690, PPARC under 
GR/L56343, NATO under CRG/94259 and by NSF.  Simulations were 
carried out at NERSC and at EPCC. We thank the UKQCD collaboration for 
providing many of the configurations and light propagators.

S.C. acknowledges a fellowship from the Royal Society of Edinburgh.
J.Sh. would like to thank members of the theoretical physics group at 
the University of Glasgow for their hospitality during an extended visit. 
Support from a PPARC Visiting Fellowship PPA/V/S/1997/00666 is 
gratefully acknowledged.

\vspace{.1in}
\noindent

\appendix

\section{Matching coefficients for $A_0$ }

In Table X we collect matching coefficients for $A_0$, 
i.e. the  $\rho_0$, $\rho_1$, 
$\rho_2$ and the mixing coefficient $\zeta_{10}$ defined in eq.(\ref{ajlat}) 
and eq.(\ref{rho0}).   This augments numbers given in reference \cite{pert1} 
where, for the case of the full NRQCD action used in the present 
simulations, a smaller set of heavy quark masses was covered.
The coefficient $\rho_2$ in Table X includes contributions from 
$J_0^{(disc)}$, as explained after eq.(\ref{jdisc}).  
We note that in the first article of reference \cite{pert1} a different 
convention was used and $\rho_2$ there did not include $2aM_0\zeta_{A_0}$.
The new convention adopted here is more in line with 
 those employed in the second article of reference \cite{pert1} for 
matching of $A_k$, $V_0$ and $V_k$.

\begin{table}
\begin{center}
\begin{tabular}{c|ccc}
    & $\beta=5.7$ & $\beta=6.0$ & $\beta=6.2$ $\quad$\\\hline
volume & $12^3\times 24$ & $16^3\times 48$ &$24^3\times 48$ \\
 \# configs & 278   & $102\times 2_{trev}$ &  144           \\
$c_{SW}$ &  1.567 (tad. imp.) &  1.479 (tad. imp.) & 1.61 (nonpert.) \\
$a^{-1}(m_\rho)$~(GeV) & 1.116(12)($^{+56}_{-0}$) \cite{hugh} &
 1.92(7)\cite{fbplb} &  2.59($^{+6}_{-10})(^{+9}_{-0}$)
\cite{joachim,rowland,goeck}  \\
$aM_0^b$& 4.20(25)(5)($^{+0}_{-24}$) \cite{joachim}& 2.22(11) \cite{fbplb}&
1.64(5)($^{+8}_{-5}$)($^{+0}_{-7}$) \cite{joachim}  \\
$\kappa_q$& 0.1400 &0.13755 &0.1346 \\
$\kappa_s$&  0.1399(1)\cite{rowland} & 0.13755(13) \cite{fbplb} & 
0.13466(7) \cite{rowland} \\
\end{tabular}
\end{center}
\caption{Simulation details for sections I through V.  
The errors on $a^{-1}$ include statistical 
errors
  and those due to the chiral extrapolation of $m_\rho$. 
 $aM_0^b$ is the bare b-quark mass in lattice units 
determined from the $B$ (or $B_s$) meson. 
$\kappa_q$ gives the light quark mass for which results are presented here
 and 
$\kappa_s$ is the actual strange $\kappa$ for these $\beta$ values based on 
the $K$ meson. 
 At $\beta=6.0$ time reversed configurations were also used and 
results were interpolated to $\kappa_s$.
  }\label{simdet}
\end{table}

\begin{table}
\begin{center}
\begin{tabular}{r|cccccccccccccccccccc}
\multicolumn{1}{c|}{} &
\multicolumn{20}{c}{$\beta = 5.7$} \\

$aM_0$ &20.0&12.5&10.0&8.0&6.0&5.0&4.0&3.5&3.15&2.75&2.45&2.2&2.0&
1.7&1.5&1.3&1.125&1.0&0.8&0.6  \\
n & 1&1&1&1&1&2&2&2&2&3&3&3&3&4&4&5&6&6&8&10 \\
\hline
\multicolumn{1}{c|}{} &
\multicolumn{20}{c}{$\beta = 6.0$} \\
$aM_0$ & & &  10.0 && &  7.0  && & 4.0 && & 2.7 && & 2.0 && & 1.6 & &  \\
n  &&& 1 &&& 1 &&&1 &&&2 &&&2 &&&2 && \\
\hline
\multicolumn{1}{c|}{} &
\multicolumn{20}{c}{$\beta = 6.2$} \\
$aM_0$ & && 4.5 &&&&&& 2.5 &&&&&& 1.44 &&&&&  \\
n  & && 1 &&&&&&3&&&&&&3&&&&& \\
\end{tabular}
\end{center}
\caption{ Bare heavy quark masses and n-values }
\end{table}

\begin{table}
\begin{center}
\begin{tabular}{c|rrr}
Source & 
\multicolumn{1}{c}{$ \beta = 5.7$ } &
\multicolumn{1}{c}{$ \beta = 6.0$ } &
\multicolumn{1}{c}{$ \beta = 6.2$ } \\\hline
statistical &  3$\;$ & 3 $\;\;$& 2 $\quad$\\
disc. $O((a\Lambda_{QCD})^2)$ &   13$\;$    & 4 $\;\;$   & 2 $\quad$  \\
disc. $O(\alpha_s \, (a\Lambda_{QCD}))$ &5$\;$    & 2 $\;\;$ & --- $\quad$  \\
pert. $O(\alpha_s^2)$, $O(\alpha_s^2/(aM))$ &  11$\;$    &   7 $\;\;$
  &  6 $\quad$ \\
NRQCD $O((\Lambda_{QCD}/M)^2)$, $O(\alpha_s \, \Lambda_{QCD}/M)$
  & 1$\;$    &  1 $\;\;$  &  1 $\quad$ \\
$\kappa_s$  &  +4$\;$   & +4$\;\;\;$   & +4  $\quad$\\
$a^{-1}(m_\rho)$ &  $(^{+5}_{-1})$   &  4$\;\;\;$  &  3 $\quad$\\\hline
Total &  {\bf 19} & {\bf 11}$\;\;$ & {\bf 8} $\quad$\\
\end{tabular}
\end{center}
\caption{Estimates of the statistical and main systematic errors, in percent, 
in our values for $f_{B_s}$}\label{errors}
\end{table}

\begin{table}
\begin{center}
\begin{tabular}{r|lllll}
\multicolumn{1}{c|}{$aM_0$} &
\multicolumn{1}{c}{$a^{3/2}f^{(0)}\sqrt{M}$} &
\multicolumn{1}{c}{$a^{3/2}f^{(1)}\sqrt{M}$} &
\multicolumn{1}{c}{$a^{3/2}f^{(3)}\sqrt{M}$} &
\multicolumn{1}{c}{$a^{3/2}f^{(4)}\sqrt{M}$} &
\multicolumn{1}{c}{$a^{3/2}f^{(5)}\sqrt{M}$} \\
\hline
\multicolumn{1}{c|}{} &
\multicolumn{4}{c}{$\beta = 5.7$} \\
\hline
 0.600 & 0.455(7)  & -0.188(4)   & -0.229(4)    & 0.1148(20)   & -0.1463(24)   \\ 
 0.800 & 0.445(7)  & -0.1448(27) & -0.1145(21)  & 0.0583(10)   & -0.0763(12)   \\
 1.000 & 0.445(8)  & -0.1204(24) & -0.0678(13)  & 0.0348(6)    & -0.0470(8)    \\
 1.125 & 0.449(8)  & -0.1105(23) & -0.0535(11)  & 0.0267(5)    & -0.0369(7)    \\
 1.300 & 0.455(9)  & -0.0992(22) & -0.0395(9)   & 0.0193(4)    & -0.0274(6)    \\
 1.500 & 0.462(9)  & -0.0893(21) & -0.0297(7)   & 0.0141(3)    & -0.0205(4)    \\
 1.700 & 0.471(10) & -0.0820(20) & -0.0238(6)   & 0.01072(25)  & -0.0161(4)    \\
 2.000 & 0.484(11) & -0.0731(19) & -0.0179(5)   & 0.00754(19)  & -0.01170(27)  \\
 2.200 & 0.492(11) & -0.0685(18) & -0.0153(4)   & 0.00616(17)  & -0.00975(23)  \\
 2.450 & 0.503(12) & -0.0636(18) & -0.0130(4)   & 0.00490(14)  & -0.00795(19)  \\
 2.750 & 0.515(13) & -0.0587(17) & -0.0108(3)   & 0.00385(12)  & -0.00639(15)  \\
 3.150 & 0.528(14) & -0.0533(16) & -0.00874(25) & 0.00289(10)  & -0.00493(12)  \\
 3.500 & 0.541(15) & -0.0495(16) & -0.00748(23) & 0.00233(9)   & -0.00405(10)  \\
 4.000 & 0.556(17) & -0.0450(16) & -0.00614(20) & 0.00177(7)   & -0.00315(8)   \\
 5.000 & 0.583(21) & -0.0381(15) & -0.00440(16) & 0.00112(5)   & -0.00207(6)   \\
 6.000 & 0.602(25) & -0.0331(15) & -0.00331(13) & 0.00077(5)   & -0.00146(4)   \\
 8.000 & 0.64(3)   & -0.0263(15) & -0.00213(10) & 0.00043(3)   & -0.00085(3)   \\
10.000 & 0.66(4)   & -0.0218(14) & -0.00149(9)  & 0.000272(25) & -0.000555(25) \\
12.500 & 0.68(6)   & -0.0180(14) & -0.00103(7)  & 0.000171(19) & -0.000362(20) \\
20.000 & 0.73(9)   & -0.0119(13) & -0.00045(4)  & 0.000066(11) & -0.000144(11) \\
\hline
\multicolumn{1}{c|}{} &
\multicolumn{4}{c}{$\beta = 6.2$} \\
\hline
1.440   & 0.143(2) & -0.0195(3)  &&& \\
2.500   & 0.147(2) & -0.0125(2)  &&& \\
4.500   & 0.152(1) & -0.0076(1)  &&&  \\
\end{tabular}
\end{center}
\caption{Matrix elements of individual current contributions 
in lattice units. 
$\;f^{(j)}$ is defined in  eq.~\protect(\ref{fsqrtmdef}). 
The $\beta=6.2$ matrix elements have different normalization from those 
at other $\beta$'s (see text).
 }
\label{treefj}
\end{table}

\begin{table}
\begin{center}
\begin{tabular}{r|llll}
\multicolumn{1}{c|}{$aM_0$} &
\multicolumn{1}{c}{tree-level} &
\multicolumn{1}{c}{$q^\ast=1/a$} &
\multicolumn{1}{c}{$q^\ast=2/a$} &
\multicolumn{1}{c}{$q^\ast=\pi/a$} \\
\hline
\hline
\multicolumn{1}{c|}{} &
\multicolumn{4}{c}{$\beta = 5.7$ $ \qquad \qquad$} \\
\hline
 0.600 &  0.315(6)  & 0.263(5)  & 0.280(5)  & 0.286(5)  \\
 0.800 &  0.353(6)  & 0.280(5)  & 0.304(5)  & 0.312(5)  \\
 1.000 &  0.382(7)  & 0.302(5)  & 0.328(6)  & 0.337(6)  \\
 1.125 &  0.399(7)  & 0.314(6)  & 0.341(6)  & 0.351(6)  \\
 1.300 &  0.419(8)  & 0.329(6)  & 0.358(7)  & 0.369(7)  \\
 1.500 &  0.440(9)  & 0.345(7)  & 0.375(8)  & 0.386(8)  \\
 1.700 &  0.459(10) & 0.364(8)  & 0.394(9)  & 0.406(9)  \\
 2.000 &  0.484(11) & 0.378(9)  & 0.412(10) & 0.424(10) \\
 2.200 &  0.500(11) & 0.385(9)  & 0.422(10) & 0.435(10) \\
 2.450 &  0.518(12) & 0.396(10) & 0.435(10) & 0.449(11) \\
 2.750 &  0.538(14) & 0.408(10) & 0.450(11) & 0.465(12) \\
 3.150 &  0.560(15) & 0.421(12) & 0.466(13) & 0.482(13) \\
 3.500 &  0.579(17) & 0.432(13) & 0.479(14) & 0.496(14) \\
 4.000 &  0.603(19) & 0.446(14) & 0.496(16) & 0.514(16) \\
 5.000 &  0.642(24) & 0.471(18) & 0.526(19) & 0.546(20) \\
 6.000 &  0.671(28) & 0.488(21) & 0.547(23) & 0.568(24) \\
 8.000 &  0.72(4)   & 0.519(28) & 0.58(3)   & 0.61(3)   \\ 
10.000 &  0.75(5)   & 0.55(4)   & 0.61(4)   & 0.64(4)   \\ 
12.500 &  0.79(7)   & 0.57(5)   & 0.64(6)   & 0.67(6)   \\ 
20.000 &  0.84(10)  & 0.63(9)   & 0.70(10)  & 0.72(10)  \\
\hline
\multicolumn{1}{c|}{} &
\multicolumn{4}{c}{$\beta = 6.0$ $\qquad \qquad$} \\
\hline
1.600&0.442(13)&0.398(11)&0.410(11)&0.414(11) \\
2.000&0.466(13)&0.406(13)&0.421(13)&0.427(13) \\
2.700&0.495(19)&0.421(16)&0.440(16)&0.448(16) \\
4.000&0.521(19)&0.442(16)&0.465(16)&0.473(16) \\
7.000&0.588(16)&0.479(13)&0.508(13)&0.518(13) \\
10.000&0.615(16)&0.498(13)&0.528(13)&0.540(13) \\
\hline
\multicolumn{1}{c|}{} &
\multicolumn{4}{c}{$\beta = 6.2$ $\qquad \qquad$} \\
\hline
1.440   &  0.515(8)   &  0.404(4)    &   0.429(8) & 0.442(8) \\
2.500   &  0.561(8)   &  0.465(6)    &   0.485(6) & 0.495(6) \\
4.500   &  0.602(4)   &  0.505(5)    &   0.527(5) & 0.536(5)\\
\end{tabular}
\end{center}
\caption{ Decay matrix elements $f\protect\sqrt{M}$ in $GeV^{3/2}$. The first
 column 
gives tree-level results, 
the remaining columns include renormalization constants using 
$\alpha_P$ at three values of $q^\ast$. Only statistical errors are shown.}
\label{tab:renorm}
\end{table}

\begin{table}
\begin{center}
\begin{tabular}{l|lcl}
\multicolumn{1}{c|}{$\beta$} &
\multicolumn{3}{c}{$f_{B_s}$ in MeV} \\
\hline
 5.7 & 217(7)(30)(24)(2)(9)($^{+12}_{\; -3}$) & $\rightarrow$ & 
217($^{+42}_{-40}$)\\
&&& \\
 6.0 & 184(6)(8)(13)(2)(7)(7) & $\rightarrow$ & 184(19)\\
&&& \\
 6.2 & 187(4)(4)(11)(2)(7)(6) & $\rightarrow$ &  187(16)\\
\end{tabular}
\end{center}
\caption{ Quenched $f_{B_s}$ with errors listed separately and added 
in quadrature to the right.  The errors denote from left to right
statistical (plus interpolations in $M_0$), discretization, perturbative,
relativistic, fixing $\kappa_s$ and  \ainv. }
\label{tab:fbs}
\end{table}

\begin{table}
\begin{center}
\begin{tabular}{c|cc}
    &  $\beta=6.0$ & $\beta=6.2$ $\quad$\\\hline
volume &  $16^3\times 48$ &$24^3\times 48$ \\
 \# configs &  268 &  144           \\
$c_{SW}$ &    1.77 (nonpert.) & 1.61 (nonpert.) \\
$aM_0$&  2.0 & 1.44 \\
$\kappa_q$& 0.13344 &0.1346 \\
\end{tabular}
\end{center}
\caption{Simulation details for section VI.  
  }
\end{table}

\begin{table}
\begin{center}
\begin{tabular}{c|cc}
    &  $\beta=6.0$ & $\beta=6.2$ $\quad$\\\hline
$aM_{pert}$   &   2.53(8)  &  1.88(5)   \\
\hline
$aM_{kin}$   &&  \\
(0,0,1) &   2.51(23)  & 1.93(33)  \\
(0,1,1) &   2.83(36)  & 1.77(19)  \\
(1,1,1) &    -----    & 1.73(17)  \\
(0,0,2) &   2.49(11)  & 1.63(16)  \\
\end{tabular}
\end{center}
\caption{ $aM_{kin}$ extracted from pseudoscalar mesons with 
different momenta and comparison with $aM_{pert}$.  Momenta are given 
in units of $2\pi/aL$.  Values for 
the lowest momenta in lattice units are: $a p =$
 0.39 ($\beta=6.0$) and 0.26 ($\beta=6.2$).  
}
\end{table}

\begin{table}
\begin{center}
\begin{tabular}{c|cc|cc}
\multicolumn{1}{c|}{} &
\multicolumn{2}{c}{$ \beta = 6.0$ $\qquad \qquad$} &
\multicolumn{2}{c}{$ \beta = 6.2$ $\qquad \qquad$} \\
\hline
momentum &  $R(p)$ & $R^{(0)}(p)$
 & $R(p)$ & $R^{(0)}(p)$  \\
\hline
(0,0,1) &  1.02(2)  & 1.01(2) & 1.01(2) & 1.00(2) \\
(0,1,1) &  1.00(5)  & 0.98(5) & 1.03(3) & 1.02(3)\\
(1,1,1) &   -----   & -----   & 1.05(4) & 1.03(4)\\
(0,0,2) &  1.08(3)  &1.05(3)  & 1.10(6) & 1.07(6)\\
\end{tabular}
\end{center}
\caption{ The ratios $R(p)$ and $R^{(0)}(p)$ of eq.(\ref{ratiof}) and 
eq.(\ref{ratio0}) for several momenta and $\beta$ values. 
  Momenta are given 
in units of $2\pi/aL$.}
\end{table}

\begin{table}
\begin{center}
\begin{tabular}{rrrrcl}
$aM_0$  &  n  &  $\rho_0 \quad$  & $\rho_1 \quad$ & 
 $ \quad \rho_2$  &$\quad \quad \zeta_{10} $  \\
\hline
0.600  &  10 & 0.1875(6)  & 0.705(3)  & 0.487(2) & -0.9944(3)\\
0.800  &   8 & -0.1191(4) & 0.213(2)  & 0.758(2) & -0.7968(2)\\
0.800  &   5 & -0.1203(4) & 0.261(2)  & 0.680(2) & -0.7682(2)\\
1.000  &   6 & -0.2207(3) & 0.026(2)  & 0.914(2) & -0.6645(1)\\
1.000  &   4 & -0.2108(3) & 0.045(2)  & 0.849(2) & -0.6502(1)\\
1.125  &   6 & -0.2546(3) &-0.035(2)  & 1.026(2) & -0.6063(1)\\
1.200  &   3 & -0.2500(3) &-0.031(2)  & 0.987(3) & -0.5613(1)\\
1.300  &  5  &  -0.2814(3) &  -0.068(2) &  1.156(3)  & -0.5387(1)\\
1.400  &  3  &  -0.2806(3) &  -0.070(2) &  1.187(3)  & -0.5014(1)\\
1.440  &  3  &  -0.2855(3) &  -0.071(2) &  1.227(3)  & -0.4909(1)\\
1.500  &  4  &  -0.2987(3) &  -0.080(2) &  1.324(3)  & -0.4790(1)\\
1.500  &  2  &  -0.2732(3) &  -0.061(2) &  1.193(3)  & -0.4675(1)\\
1.600  &  2  &  -0.2845(2) &  -0.065(2) &  1.311(4)  & -0.4463(1)\\
1.700  &  2  &  -0.2941(2) &  -0.064(2) &  1.425(4)  & -0.4267(1)\\
2.000  &  2  & -0.3146(2)  &  -0.046(2) &  1.781(4)  & -0.3768(1)\\
2.500  &  3  &  -0.3382(2) &  -0.006(2) &  2.454(6)  & -0.31626(7)\\
2.500  &  2  &  -0.3331(2) &  -0.001(2) &  2.432(6)  & -0.31525(7)\\
2.700  &  2  &  -0.3374(2)  &   0.018(2)&  2.706(6)  & -0.29599(6)\\
3.000  &  2  &  -0.3421(2)  &   0.044(2)&  3.132(7)  & -0.27122(6)\\
3.500  &  2  &  -0.3457(2) &   0.094(2) &  3.895(8)  & -0.23818(6)\\
4.000  &  2  &  -0.3460(2)  &   0.135(2)&  4.705(10) & -0.21235(5)\\
4.000  &  1  &  -0.3374(2)  &   0.140(2)&  4.600(10)  & -0.21147(5)\\
4.500  &  1  &  -0.3379(2)  &   0.176(3)&  5.467(11) & -0.19115(5)\\
5.000  &  1  &  -0.3362(2)  &  0.211(3) &  6.328(12) & -0.17433(4)\\
7.000  &  1  &  -0.3173(2)  &   0.321(4)& 10.050(21) & -0.12898(3)\\
10.000 &  1  &  -0.2770(2)  &  0.432(6) & 15.957(30) & -0.09282(3)\\
12.500  &  1  &  -0.2432(2)  &  0.505(7)& 20.905(36) & -0.07523(2)\\
20.000  &  1  &  -0.1528(2)  &   0.655(12)& 36.335(64) & -0.04797(1)\\
\end{tabular}
\end{center}
\caption{ One-loop matching coefficients $\rho_0$, $\rho_1$ and 
$\rho_2$ of eq.(\ref{ajlat}) and mixing coefficient $\zeta_{10}$ 
of eq.(\ref{rho0}) for several heavy quark mass values. 
 $\rho_2$ includes 
$ 2 aM_0\, \zeta_{A_0}$ in order to incorporate contributions from 
$J^{(disc)}$.  The NRQCD action is that of eq.(\ref{nrqcdact}) - 
(\ref{deltaH}) with $c_i = 1$.  Both the NRQCD and clover light quark actions 
are taken to be tadpole-improved with the plaquette definition of $u_0$.
}
\end{table}

\newpage
\begin{figure}
\begin{center}
\epsfysize=7.in
\centerline{\epsfbox{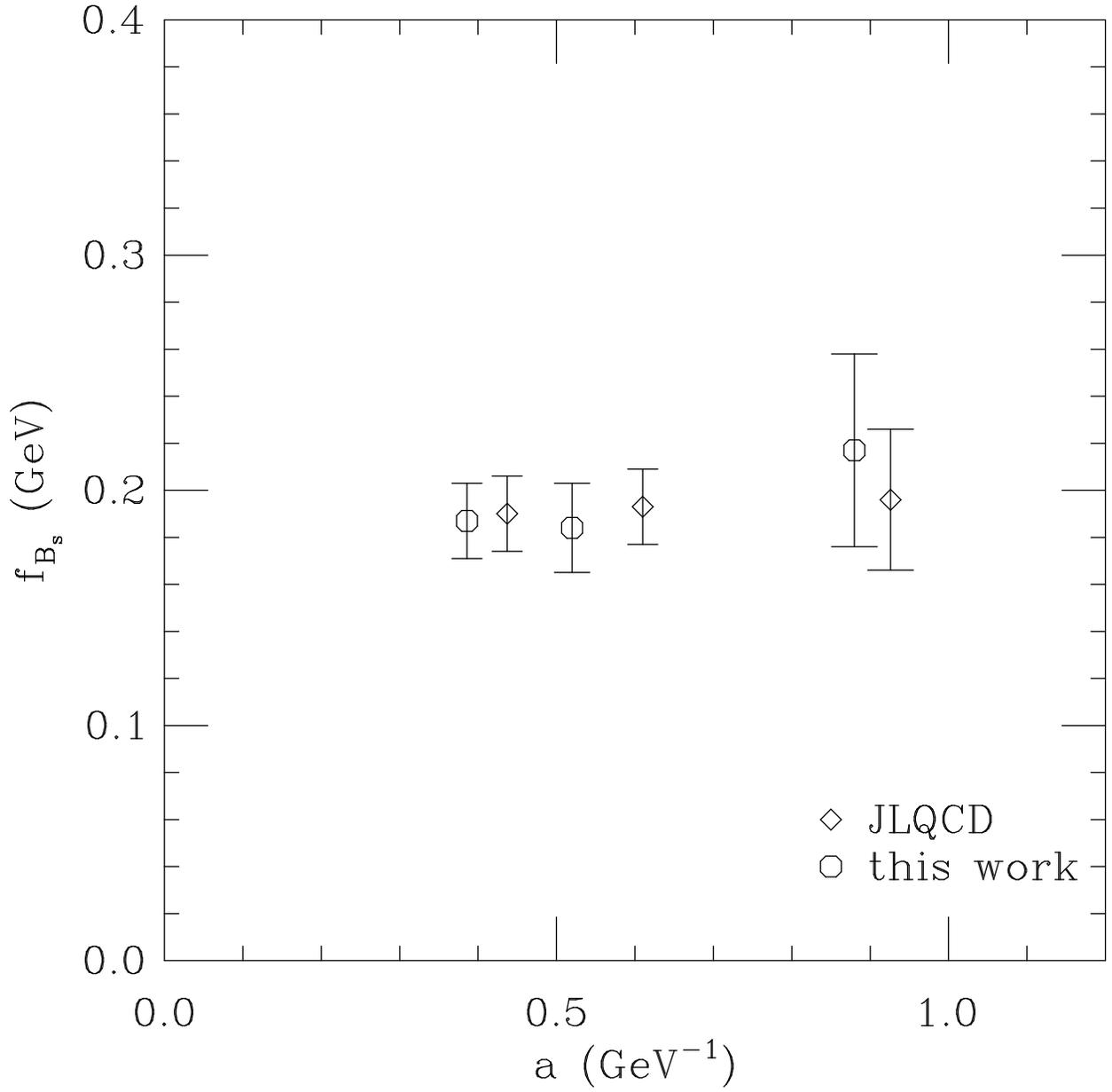 }}
\end{center}
\caption[fig:one]{
\fbs versus the lattice spacing and comparison with reference \cite{jlqcdhiro}.
 Systematic and statistical errors have been added in quadrature.
   }
\label{fig:one}
\end{figure}

\newpage
\begin{figure}
\begin{center}
\epsfysize=7.in
\centerline{\epsfbox{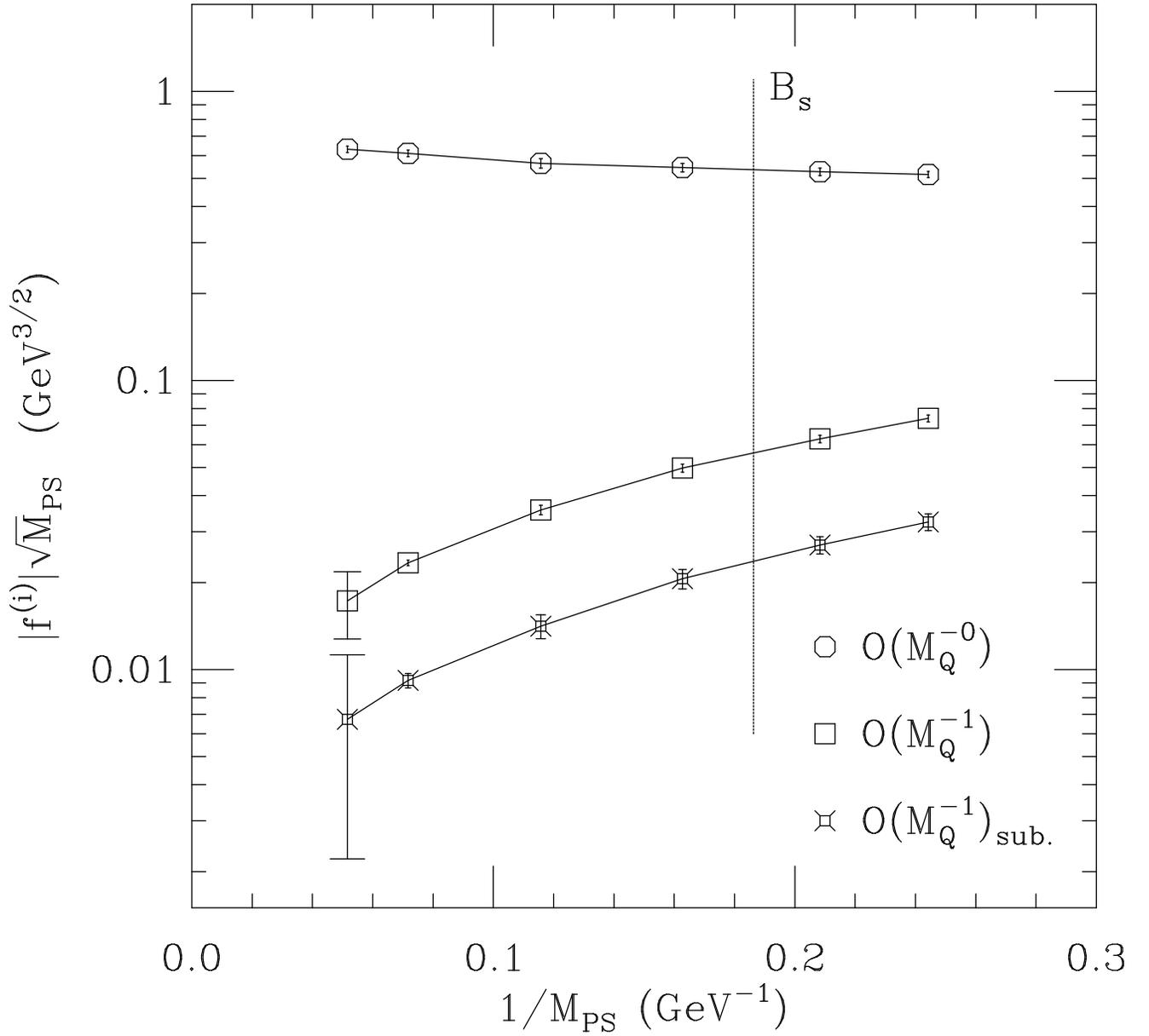 }}
\end{center}
\caption[fig:two]{
Matrix elements $f^{(0)}\sqrt{M}$, $|f^{(1)}|\sqrt{M}$ and 
$|f^{(1)}_{sub}|\sqrt{M}$ versus $1/M_{PS}$ at $\beta = 6.0$.  
$q^\ast = 2/a$ was used to obtain 
$|f^{(1)}_{sub}|\sqrt{M}$.  
  }
\label{fig:two}
\end{figure}

\newpage
\begin{figure}
\begin{center}
\epsfysize=7.in
\centerline{\epsfbox{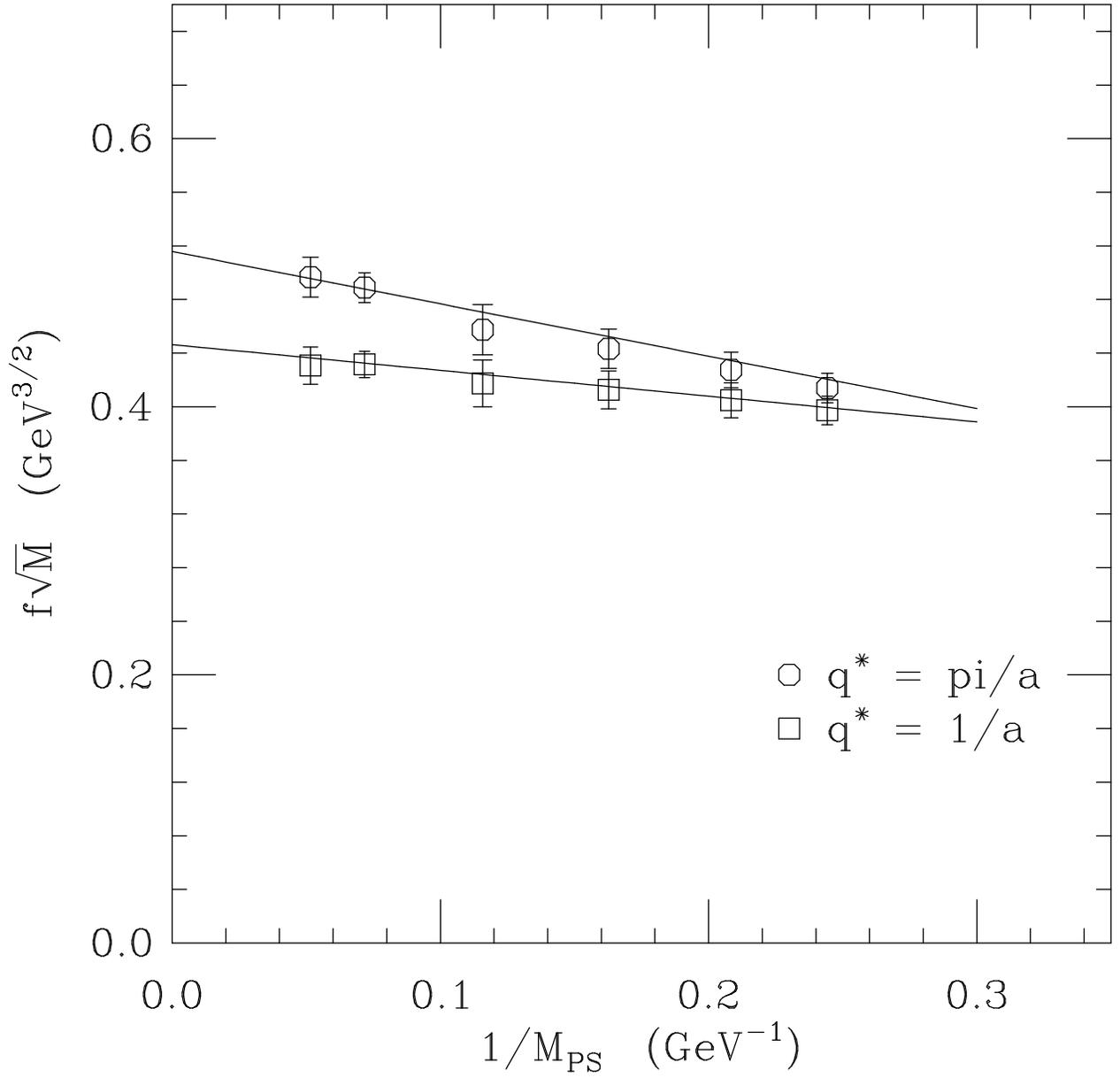 }}
\end{center}
\caption[fig:three]{
$f_{PS}\sqrt{M_{PS}}$ versus $1/M_{PS}$ 
at $\beta = 6.0$ for $q^\ast=\pi/a$ and $q^\ast = 1/a$.
The lines are linear fits to the data.  Only statistical errors are shown.
  }
\label{fig:three}
\end{figure}

\newpage
\begin{figure}
\begin{center}
\epsfysize=7.in
\centerline{\epsfbox{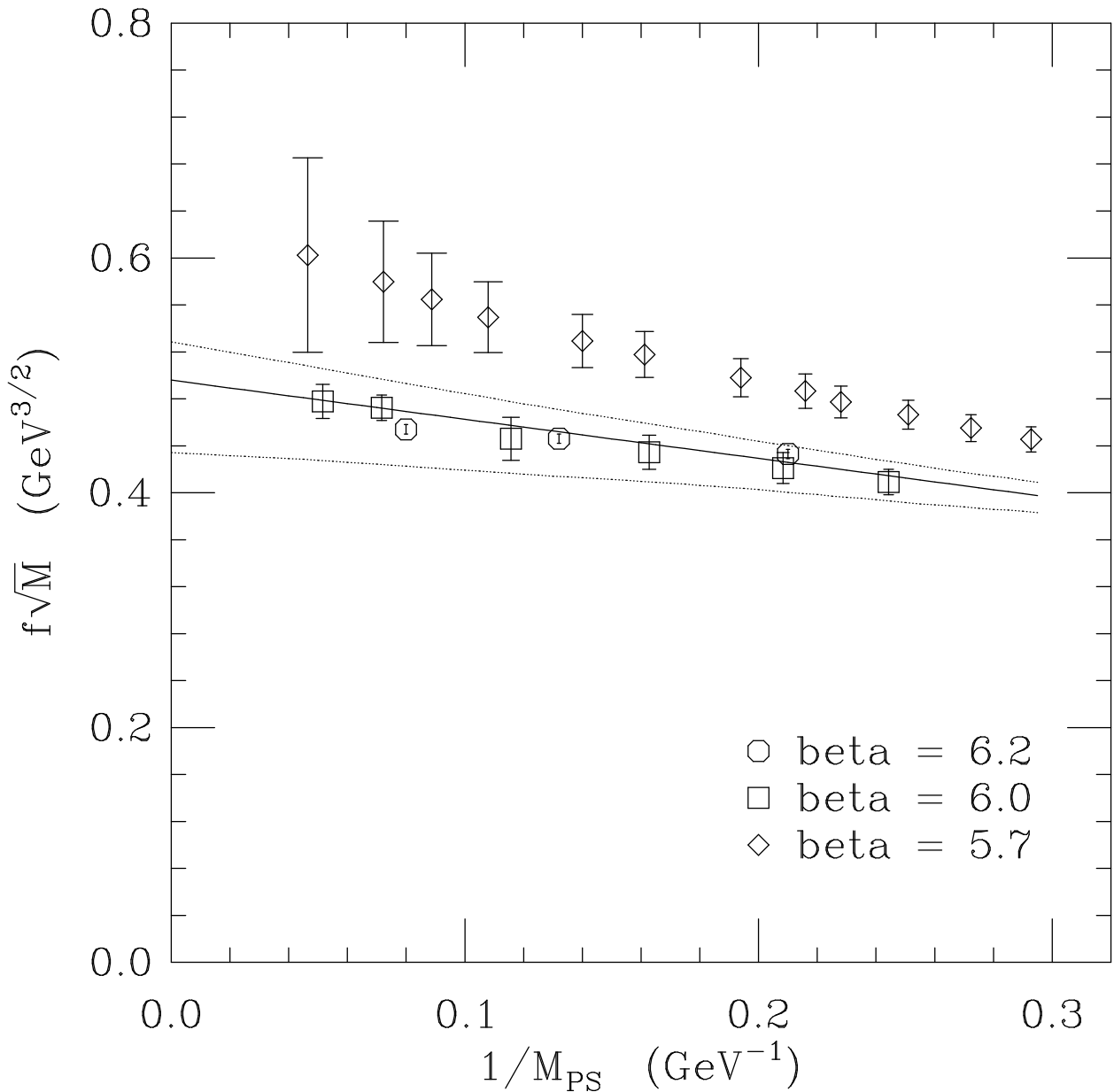 }}
\end{center}
\caption[fig:four]{
Results for $f_{PS}\sqrt{M_{PS}}$ versus $1/M_{PS}$ from different $\beta$'s. 
$q^\ast = 2/a$ is used for the one-loop renormalized data points.
The central full
 line is a linear fit to the $q^\ast = 2/a$, $\beta = 6.0$ data. 
The bottom(top) dotted line comes from lower(upper) bounds on fits to 
$q^\ast=1/a$($q^\ast=\pi/a$), $\beta=6.0$ data (see text).
  }
\label{fig:four}
\end{figure}

\newpage
\begin{figure}
\begin{center}
\epsfysize=7.in
\centerline{\epsfbox{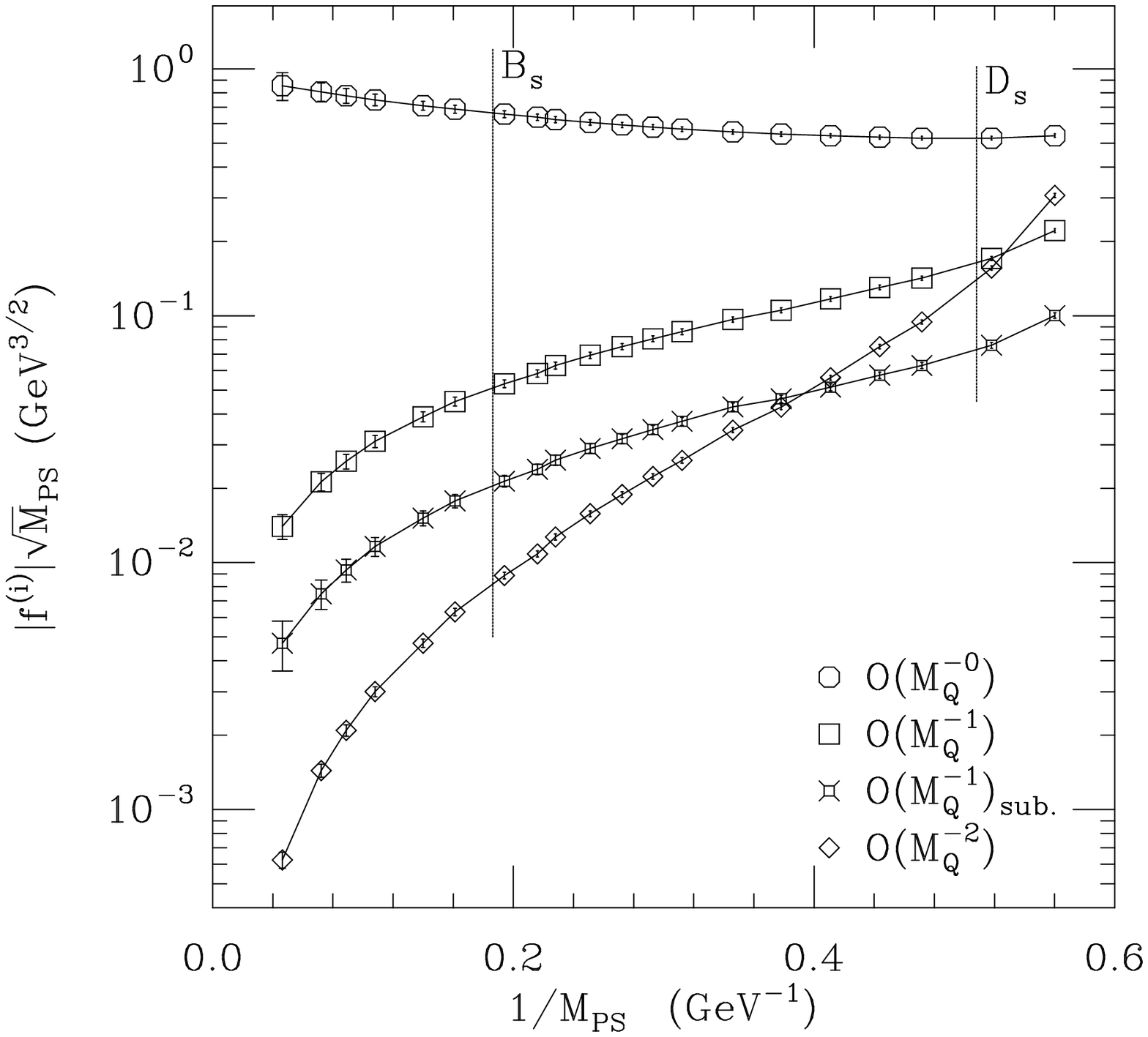 }}
\end{center}
\caption[fig:five]{
Matrix elements $f^{(0)}\sqrt{M}$, $|f^{(1)}|\sqrt{M}$, 
 $|f^{(1)}_{sub}|\sqrt{M}$ and 
$(|f^{(3)} + f^{(4)} + f^{(5)}|)\sqrt{M}$ versus $1/M_{PS}$ at $\beta = 5.7$.  
$q^\ast = 2/a$ was used to obtain 
$|f^{(1)}_{sub}|\sqrt{M}$.  
  }
\label{fig:five}
\end{figure}

\newpage
\begin{figure}
\begin{center}
\epsfysize=7.in
\centerline{\epsfbox{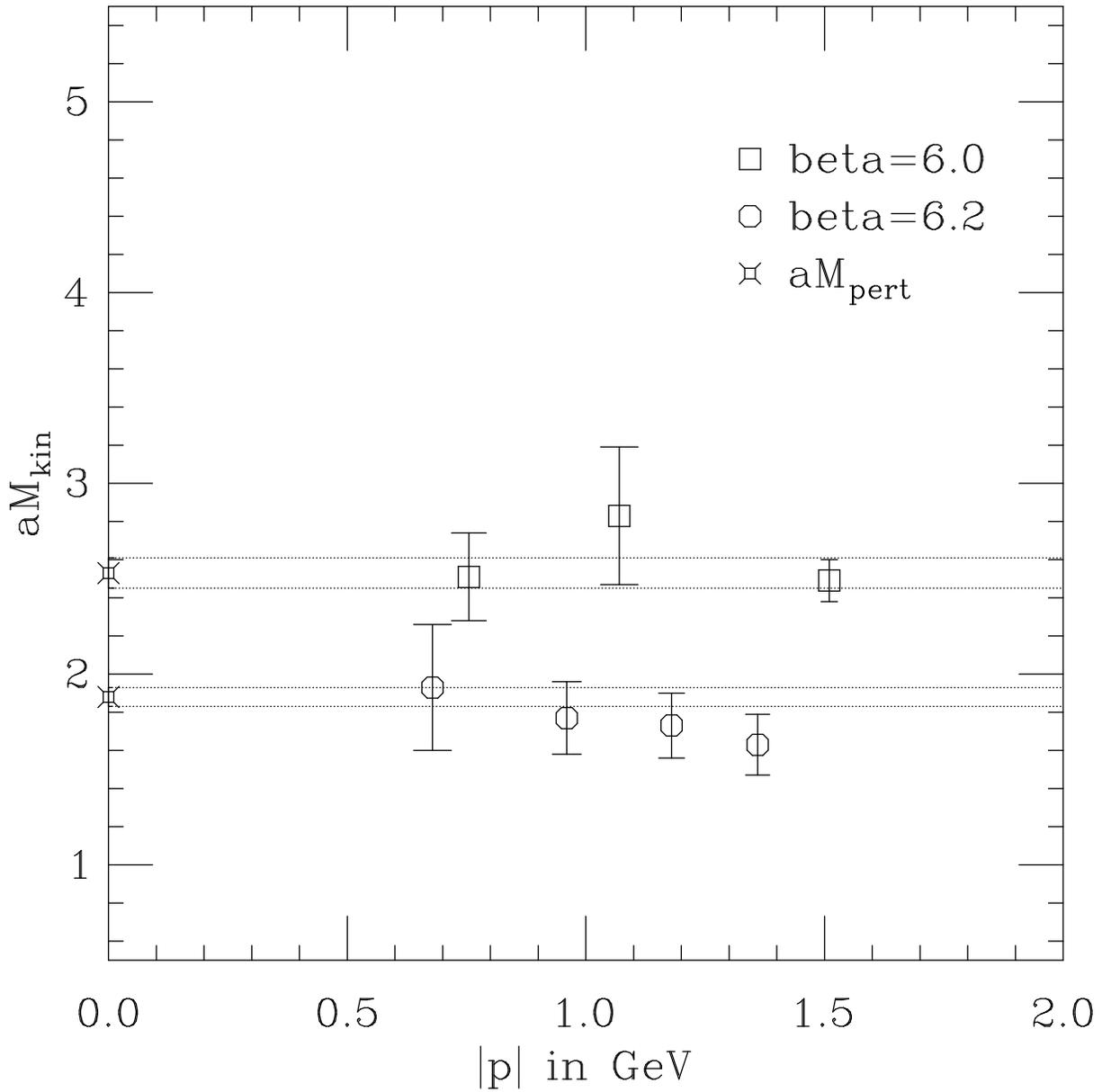 }}
\end{center}
\caption[fig:six]{
$aM_{kin}$ extracted from mesons with different spatial momenta.  The 
fancy squares to the left show $aM_{pert}$.  The horizontal lines 
bracket perturbative errors on $aM_{pert}$.
  }
\label{fig:six}
\end{figure}

\newpage
\begin{figure}
\begin{center}
\epsfysize=7.in
\centerline{\epsfbox{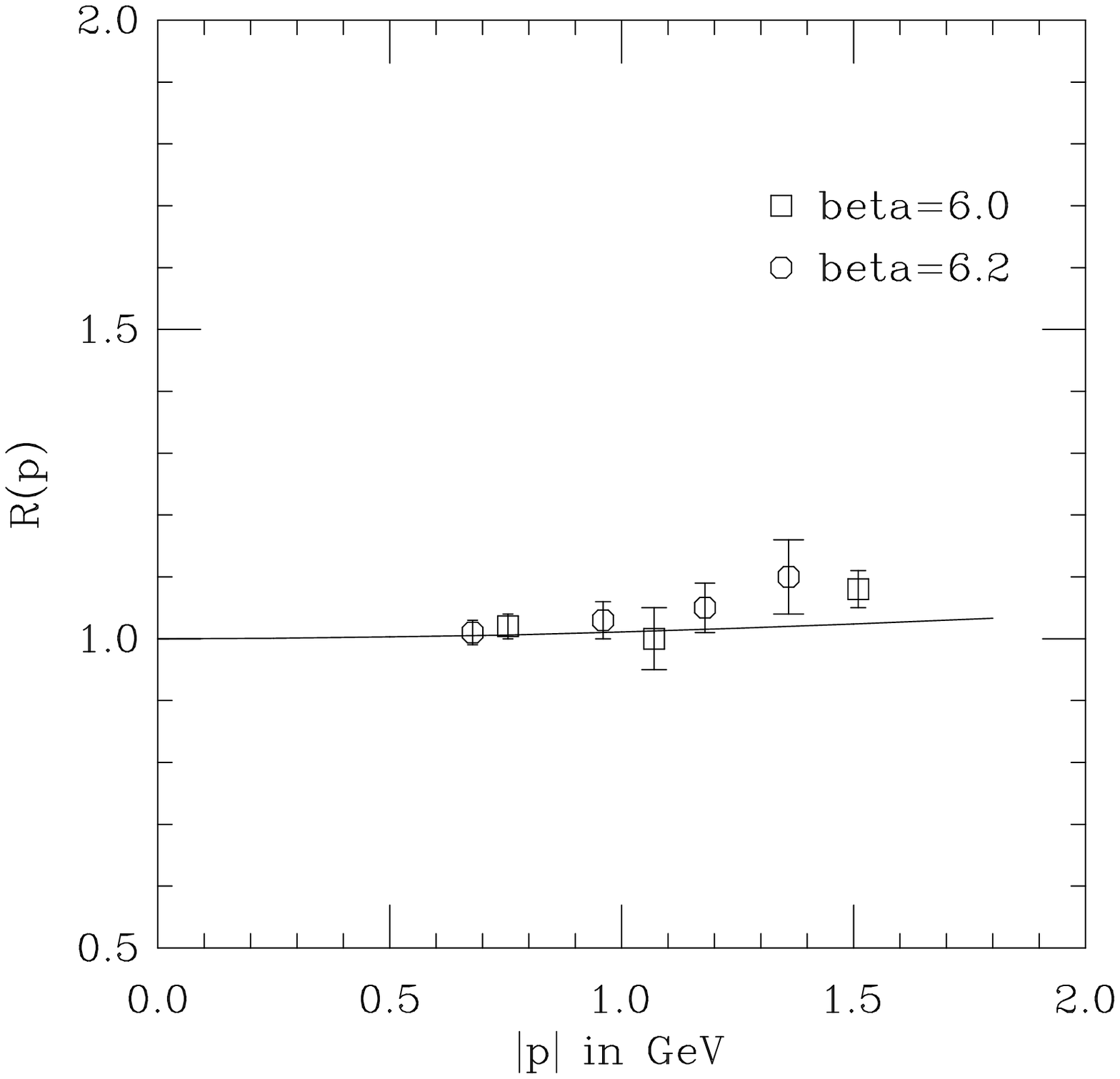 }}
\end{center}
\caption[fig:seven]{
The ratio $R(p)$ of eq.(\ref{ratiof}) versus momentum.  The full line 
shows $\sqrt{E(p)} / \sqrt{M_{PS}}$.  
  }
\label{fig:seven}
\end{figure}


\begin{thebibliography}{8}

\bibitem{fb}
S. Hashimoto, review talk given at {\it Lattice} '99,
 Nucl. Phys. Proc. Suppl. {\bf 83-84}, 3 (2000).


\bibitem{jlqcd98}
JLQCD collaboration, S. Aoki {\em et al.},
 Phys. Rev. Lett. {\bf80}, 5711 (1998).

\bibitem{fbplb}
A. Ali Khan {\em et al.}, Phys. Lett. {\bf B427}, 132 (1998). 

\bibitem {jlqcdhiro}
JLQCD collaboration, K-I. Ishikawa {\em et al.}, Phys. Rev. 
{\bf D61}:074501 (2000).

\bibitem{fermifb}
A.X. El-Khadra {\em et al.}, Phys. Rev. {\bf D58}:014506 (1998).

\bibitem{milc}
MILC collaboration, C. Bernard {\em et al.}, Phys. Rev. Lett. {\bf 81}, 4812
 (1998).

\bibitem{ape}
APE collaboration, D. Becirevic {\em et al.}, Phys. Rev. 
{\bf D60}:074501 (1999).

\bibitem{ukqcd}
UKQCD collaboration, C. Maynard {\em et al.}, 
 Nucl. Phys. Proc. Suppl. {\bf 83-84}, 322 (2000).


\bibitem{sara}
S. Collins {\em et al.}, Phys. Rev. {\bf D60}:074504 (1999).

\bibitem{cppacs1}
CP-PACS collaboration, A. Ali Khan {\em et al.}, 
 Nucl. Phys. Proc. Suppl. {\bf 83-84}, 265 (2000).


\bibitem{cppacs2}
CP-PACS collaboration, H. Shanahan {\em et al.}, 
 Nucl. Phys. Proc. Suppl. {\bf 83-84}, 331 (2000).


\bibitem {cornell}
G.P.Lepage {\em et al.}, 
 Phys. Rev. {\bf D46}, 4052 (1992).



\bibitem{heavy}
A. El-Khadra, A. Kronfeld and P. Mackenzie, Phys. Rev. {\bf D55}, 
3933 (1997).


\bibitem{hugh}
UKQCD collaboration, H. Shanahan {\em et al.}, Phys. Rev. {\bf D55}, 
1548 (1997).

\bibitem{joachim}
J. Hein {\em et al.}, Phys. Rev. {\bf D62}:074503 (2000).

\bibitem{rowland}
UKQCD collaboration, P.A. Rowland, PhD thesis, University of Edinburgh, 
1997.

\bibitem{goeck}
M. G\"ockeler {\em et al.}, Phys. Rev. {\bf D57}, 5562 (1998).



\bibitem{jhein98}
J. Hein {\em et al.},
 Nucl. Phys. Proc. Suppl. {\bf 73}, 366 (1999).

\bibitem{sara99}
S. Collins {\em et al.}, 
 Nucl. Phys. Proc. Suppl. {\bf 83-84}, 271 (2000).


\bibitem {pert1}
C. Morningstar and J. Shigemitsu; Phys. Rev. {\bf D57}, 6741 (1998),
  Phys. Rev. {\bf D59}:094504 (1999).


\bibitem{fbfromak}
S. Collins {\em et al.}, in preparation



\bibitem {lepmac}
 G.P. Lepage and  P.B. Mackenzie, Phys. Rev.  {\bf D48}, 2250 (1993).


\bibitem {alpha}
K. Jansen {\em et al.}, Phys. Lett. B{\bf 372}, 275 (1996).

\bibitem{nrqcdalpha}
C.T.H. Davies {\em et al.}, Phys. Lett. {\bf B345}, 42 (1995). 

\bibitem{blm}
S.J. Brodsky, G.P. Lepage and P.B. Mackenzie, Phys. Rev. {\bf D28}, 228 (1983).

\bibitem {herhill}
O. Hernandez and B. Hill, Phys. Rev {\bf D50}, 495 (1994).

\bibitem{craige}
C. Bernard, M. Golterman and C. McNeile, Phys. Rev. {\bf D59}:074506 (1999).

\bibitem{crisa}
M. Crisafulli, V. Lubicz and A. Vladikas, Eur. Phys. J. {\bf C4}, 145 (1998).

\bibitem{draper}
T. Draper, review talk given at {\it Lattice} '98, 
 Nucl. Phys. Proc. Suppl. {\bf 73}, 43 (1999).

\bibitem{expfds}
F. Parodi, P. Roudeau and A. Stocchi, Nuovo Cim. {\bf A112}, 833 (1999).

\bibitem{simone}
Similar studies are reported in : J.N. Simone, 
 Nucl. Phys. Proc. Suppl. {\bf 53}, 386 (1997).

 \end{thebibliography}
\end{document}